\documentclass[]{aastex62}

\usepackage{xspace}

\newcommand{\update}{}
\newcommand{\starname}{LTT~9779} 
\newcommand{\planet}{LTT~9779b} 
\newcommand{\tirr}{\ensuremath{T_\mathrm{irr}}}
\newcommand{\TESS}{{\em TESS}} 
\newcommand{\transittimeone}{2458781.13997}  
\newcommand{\utransittimeone}{0.00032}
\newcommand{\transittimetwo}{2458783.51684}  
\newcommand{\utransittimetwo}{0.00053}
\newcommand{\transittime}{2458783.51636}  
\newcommand{\utransittime}{0.00027}
\newcommand{\rprsone}{0.0431}  
\newcommand{\urprsone}{0.0011}  
\newcommand{\rprstwo}{0.0452}  
\newcommand{\urprstwo}{0.0017}  
\newcommand{\newperiod}{0.79207022}
\newcommand{\unewperiod}{0.00000069}
\newcommand{\phaseamplitude}{358}
\newcommand{\uphaseamplitude}{106}
\newcommand{\dayflux}{375}
\newcommand{\udayflux}{62}
\newcommand{\nightflux}{17}
\newcommand{\unightflux}{123}

\newcommand{\phaseoffset}{\ensuremath{-10}} 
\newcommand{\eastphaseoffset}{\ensuremath{-10}} 
\newcommand{\uphaseoffset}{\ensuremath{21}}
\newcommand{\thot}{1800}
\newcommand{\uthot}{120}
\newcommand{\tcold}{700}
\newcommand{\utcold}{430}
\newcommand{\deltat}{1110}
\newcommand{\udeltat}{460} 
\newcommand{\utwodeltat}{720} 
\newcommand{\uthreedeltat}{1030} 
\newcommand{\ab}{0.72}
\newcommand{\uab}{0.12}
\newcommand{\recirc}{0.06}  
\newcommand{\urecirc}{0.18}
\newcommand{\uppertworecirc}{0.49}

\newcommand{\abtwo}{0.71}

\newcommand{\uabtwo}{0.04}
\newcommand{\radadvupperlimit}{3.8}

\shorttitle{Phase Curve of a High-Metallicity Hot Neptune}
\shortauthors{Crossfield et al.}

\begin{document}

\title{Phase Curves of Hot Neptune \planet\ Suggest a High-Metallicity
  Atmosphere}

\correspondingauthor{Ian J.\ M.\ Crossfield}
\email{ianc@ku.edu}

\author{Ian J.\ M.\ Crossfield}
\affiliation{Department of Physics and Astronomy, University of
  Kansas, Lawrence, KS, USA}

\author{Diana Dragomir}
\affiliation{Department of Physics and Astronomy, University of New Mexico, Albuquerque, NM, USA}

\author[0000-0001-6129-5699]{Nicolas B. Cowan}
\affiliation{Department of Physics, McGill University, Montr\'{e}al,
QC H3A 2T8, Canada}
\affiliation{Department of Earth \& Planetary Sciences, McGill
University, Montr\'{e}al, QC H3A 2T8, Canada}

\author[0000-0002-6939-9211]{Tansu Daylan}
\affiliation{Department of Physics and Kavli Institute for Astrophysics and Space Research, Massachusetts Institute of Technology, Cambridge, MA, USA}
\affiliation{MIT Kavli Fellow}

\author[0000-0001-9665-8429]{Ian~Wong}
\affiliation{Department of Earth, Atmospheric and Planetary Sciences, Massachusetts Institute of Technology, Cambridge, MA 02139, USA}
\affiliation{51 Pegasi b Fellow}

\author{Tiffany Kataria}
\affiliation{NASA Jet Propulsion Laboratory, 4800 Oak Grove Drive, Pasadena, CA, USA}

\author{Drake Deming}
\affiliation{Department of Astronomy, University of Maryland, College Park, MD, USA}

\author{Laura Kreidberg}
\affiliation{Max-Planck Institut f\"ur Astronomie, K\"onigstuhl 17, 69117, Heidelberg, Germany}

\author{Thomas Mikal-Evans}
\affiliation{Department of Physics and Kavli Institute for Astrophysics and Space Research, Massachusetts Institute of Technology, Cambridge, MA, USA}

\author{Varoujan Gorjian}
\affiliation{Jet Propulsion Laboratory, California Institute of Technology, Pasadena, CA, USA}

\author{James S.\ Jenkins}
\affiliation{Departamento de Astronomia, Universidad de Chile, Camino del Observatorio 1515, Las Condes, Santiago, Chile}
 \affiliation{Centro de Astrof\'isica y Tecnolog\'ias Afines (CATA), Casilla 36-D, Santiago, Chile}


\author{Bj\"orn Benneke}
\affiliation{Departement de Physique, and Institute for Research on Exoplanets, Universite de Montreal, Montreal, Canada}

\author[0000-0001-6588-9574]{Karen A.\ Collins}
\affiliation{Center for Astrophysics \textbar \ Harvard \& Smithsonian, 60 Garden Street, Cambridge, MA, USA}

\author[0000-0002-7754-9486]{Christopher~J.~Burke}
\affiliation{Department of Physics and Kavli Institute for Astrophysics and Space Research, Massachusetts Institute of Technology, Cambridge, MA, USA}

\author{Christopher E. Henze}
  \affiliation{NASA Ames Research Center, Moffett Field, CA, USA}


\author{Scott McDermott} 
\affiliation{Proto-Logic LLC, 1718 Euclid Street NW, Washington, DC 20009, USA}

\author[0000-0002-4510-2268]{Ismael Mireles}
\affiliation{Department of Physics and Kavli Institute for Astrophysics and Space Research, Massachusetts Institute of Technology, Cambridge, MA, USA}


\author[0000-0002-3555-8464]{David Watanabe} 
\affiliation{Planetary Discoveries in Fredericksburg VA 22405}

  \author[0000-0002-5402-9613]{Bill Wohler}
  \affiliation{NASA Ames Research Center, Moffett Field, CA, 94035, USA}

  \author{George Ricker}
\affiliation{Department of Physics and Kavli Institute for Astrophysics and Space Research, Massachusetts Institute of Technology, Cambridge, MA, USA}

\author{Roland Vanderspek}
\affiliation{Department of Physics and Kavli Institute for Astrophysics and Space Research, Massachusetts Institute of Technology, Cambridge, MA, USA}


 \author[0000-0002-6892-6948]{Sara~Seager}
\affiliation{Department of Physics and Kavli Institute for Astrophysics and Space Research, Massachusetts Institute of Technology, Cambridge, MA, USA}
\affiliation{Department of Earth, Atmospheric and Planetary Sciences, Massachusetts Institute of Technology, Cambridge, MA 02139, USA}
\affiliation{Department of Aeronautics and Astronautics, MIT, 77 Massachusetts Avenue, Cambridge, MA 02139, USA}


\author{Jon M.\ Jenkins}
\affiliation{NASA Ames Research Center, Moffett Field, CA, USA}



\begin{abstract}
Phase curve measurements provide a global view of the composition,
thermal structure, and dynamics of exoplanet atmospheres. Although
most of the dozens of phase curve measurements made to date are of
large, massive hot Jupiters, there is considerable interest in probing
the atmospheres of the smaller planets that are the more typical end
product of the planet formation process. One such planet that is
favorable for these studies is the ultra-hot Neptune LTT~9779b,
{\update a rare denizen of the Neptune desert}. A companion paper
presents the planet's secondary eclipses and day-side thermal emission
spectrum; in this work we describe the planet's optical and infrared
phase curves, characterized using a combination of {\em Spitzer} and
\TESS\ photometry.  We detect \starname b's thermal phase variations
at 4.5\,\micron, finding a phase amplitude of $\phaseamplitude \pm
\uphaseamplitude$~ppm and no significant phase offset, with a
longitude of peak emission occurring $\phaseoffset^\mathrm{o} \pm
\uphaseoffset^\mathrm{o}$ {\update east} of the substellar
point. Combined with our secondary eclipse observations, these phase
curve measurements imply a 4.5\,\micron\ day-side brightness
temperature of $\thot \pm \uthot$~K, a night-side brightness
temperature of $\tcold \pm \utcold$~K ($<1350$~K at $2\sigma$
confidence), and a day-night brightness temperature contrast of
$\deltat \pm \udeltat$~K. We compare our data to the predictions of 3D
general circulation models calculated at multiple metallicity levels
and to similar observations of hot Jupiters experiencing similar
levels of stellar irradiation.  {\update Though not conclusive, our
  measurement of its small 4.5\,\micron\ phase offset, the relatively
  large amplitude of the phase variation, and the qualitative
  differences between our target's day-side emission spectrum and
  those of hot Jupiters of similar temperatures all suggest} a
super-Solar atmospheric metallicity for LTT~9779b, {\update as might
  be expected given its size and mass}.  Finally, we
measure the planet's transits at both 3.6\,\micron\ and 4.5\,\micron,
providing a refined ephemeris ($P=\newperiod\pm\unewperiod$~d,
$T_0=\transittime\pm\utransittime$, BJD$_\mathrm{TDB}$) that will
enable efficient scheduling of future observations to further
characterize the atmosphere of this intriguing planet.


\end{abstract}


\section{Introduction}
\label{sec:intro}

Planets are inherently three-dimensional objects, with variation in
temperature, chemistry, and cloud coverage throughout their
atmospheres. By monitoring the brightness of a transiting exoplanet
system over the course of an entire orbital period we measure its
phase curve --- how the planet's brightness varies at different
viewing angles. Thermal phase curve observations, which record the
changes in a planet's observed brightness at infrared wavelengths, are
a powerful technique to reveal the 3D nature of exoplanet atmospheres.
Phase curve observations provide a wealth of information about
planetary atmospheric dynamics and energetics by measuring
longitudinal brightness temperature maps, thereby constraining
atmospheric conditions across the planet's surface \citep[][and
  references therein]{heng:2015,parmentier:2018hb}.

In particular, two key observables provide these insights. First, the
phase offset \citep[equal to the longitude of peak brightness in the
  simple models we consider here; see][]{schwartz:2017} indicates the
efficiency with which winds circulate incident stellar energy around
the planet.  A nonzero phase offset suggests heat transport around the
planet, while a ~zero phase offset implies the incident stellar flux
is re-radiated or inhibited by other means \citep[e.g., magnetic
  drag;][]{menou:2012}.  Second, the phase curve amplitude gives the
day-to-night temperature contrast, with low amplitudes again indicating
globally efficient redistribution of incident stellar irradiation
around the planet.  The planetary day-side emission can be inferred
from the secondary eclipse depth, which are typically observed at the
beginning and end of phase curve observations to calibrate the
baseline stellar flux. Additional parameters such as Bond albedo, heat
recirculation efficiency, and various atmospheric timescales can be
estimated based on infrared phase curves
\citep{cowan:2007,cowan:2011,cowan:2011circ}. Phase curve observations
can also be compared against predictions of 3D general circulation
models (GCMs), which can  self-consistently couple atmospheric dynamics
with radiative transfer \citep[e.g.,][]{showman:2009}.  In the most
observationally favorable systems, spectroscopic phase curves
(observed at many wavelengths simultaneously) can provide
longitudinally-averaged emission spectra, thermal profiles, and
abundances across the entire planet over a range of altitudes
\citep{stevenson:2014c,arcangeli:2019}.

Despite the many insights provided by infrared phase curves, to date
such observations have been largely limited to hot Jupiters, which are
brighter and therefore easier to characterize than the smaller planets
that occur more frequently on short-period orbits \citep{howard:2012}.
Smaller, lower-mass planets may also have qualitatively different
atmospheres than more massive hot Jupiters, e.g.\ with higher atmospheric
metallicity \citep{fortney:2013} or atmospheres further from chemical
equilibrium \citep{line:2011,moses:2013} than those of hot
Jupiters. Despite dozens of hot Jupiter phase curves, to date infrared
phase curves have been reported for just three exoplanets
substantially smaller than Jupiter (see
Table~\ref{tab:smallphasecurves}): GJ~436b \citep{stevenson:2012b},
55~Cnc~e \citep{demory:2016b}, and LHS~3844b \citep{kreidberg:2019}.


\hspace{-1in}
\begin{deluxetable}{l c r c r l}[bt]
\hspace{-1in}\tabletypesize{\scriptsize}
\tablecaption{  Non-Giant Exoplanets with Thermal Phase Curves \label{tab:smallphasecurves}}
\tablewidth{0pt}
\tablehead{
  &&&&\\
  \colhead{} & \colhead{Wavelength}    & \colhead{Radius} & \colhead{Density} & \colhead{Irradiation} & \colhead{} \\
\colhead{Planet} & \colhead{[\micron]} & \colhead{[$R_\oplus$]} & \colhead{[g~cm$^{-3}$]}  & \colhead{[$S_\oplus$]} & \colhead{References}
}
\startdata
LTT 9779 b & 4.5 & 4.72$\pm$0.23   & 1.53$\pm$0.13 & 2420 $\pm$ 140 & \cite{jenkins:2020}; This Work \\
GJ 436 b   & 8.0 & 4.04$\pm$0.17   & 2.11$\pm$0.33 & 28.8 $\pm$ 2.0 & \cite{stevenson:2012b,bourrier:2018a}\\
55 Cnc e   & 4.5 & 1.875$\pm$0.029 & 6.66$\pm$0.42 & 2397 $\pm$ 36  & \cite{demory:2016b,bourrier:2018b}  \\
LHS 3844 b & 4.5 & 1.303$\pm$0.022 & unknown           & 69.9 $\pm$ 7.1 & \cite{vanderspek:2019,kreidberg:2019}\\
\enddata
\end{deluxetable}

Here we present a new infrared phase curve of LTT~9779b (also known as
TOI-193b), a transiting hot Neptune recently discovered by the
\TESS\ survey \citep{guerrero:2020,jenkins:2020} and subsequently
observed by us with the {\em Spitzer} Space Telescope.  The 4.6
$R_\oplus$ planet is in a $0.8$~d orbit around its G-dwarf host star,
giving it an irradiation temperature $\tirr \equiv T_\mathrm{eff}
\sqrt{R_*/a} = 2770$~K and an equilibrium temperature (assuming zero
Bond albedo and complete heat recirculation) of $T_\mathrm{eq} =
0.25^{1/4} \tirr = 1960~K$. LTT~9779's moderate size and high
irradiation level make it a rare inhabitant of the so-called ``Neptune
Desert'' \citep{mazeh:2016} and an excellent target for studying the
emergent thermal spectrum of such a world.

This paper presents LTT~9779b's optical and infrared phase curves, and
builds on our analysis of the planet's secondary eclipses and day-side
emission spectrum  \citep{dragomir:2020}. Sec.~\ref{sec:obs}
describes {\em TESS} and {\em Spitzer}'s space-based photometry of the
system, and we discuss our analysis of these data in
Sec.~\ref{sec:lcanalysis}. Our derived measurements of the planet's
phase curves and associated transits are presented in
Sec.~\ref{sec:measure}, and we discuss the implications of these
measurements and conclude in Sec.~\ref{sec:discinterp}.

\section{Observations}
\label{sec:obs}
LTT~9779 was observed by the \TESS\ mission \citep{ricker:2016} in the
second 27-day sector of the sky to be observed in its two-year,
nearly-all-sky survey. It was observed from 2018/08/23 until
2018/09/20 at a two-minute observing cadence. The observations are
nearly continuous, except for a 32-hour break from 2018/09/05--07 for
data downlink.

After the full sector’s data was downlinked they were processed by the
Science Processing Operations Center \citep[SPOC;][]{jenkins:2016}
which identified LTT 9779b's transit signature and provided vetting
diagnostics and an initial model fit
\citep{jenkins:2002,twicken:2018,li:2019}. LTT 9779b was released as a
TESS alert on 3 November 2018 as TESS Object of Interest (TOI) 193.01
\citep[][]{guerrero:2020}.  The planet was subsquently validated via
radial velocity measurements \citep{jenkins:2020}.

Soon after the planet candidate was announced, we identified it as an
exceptional target for atmospheric characterization via secondary
eclipses and phase curves, including with the {\em Spitzer} Space
Telescope. As described by \cite{dragomir:2020}, we observed eight
eclipses (four at 3.6\,\micron\ and four at 4.5\,\micron) in {\em
  Spitzer} program GO-14084 \citep{crossfield:2018spitzer} as part of
a dedicated {\em TESS} follow-up program. We also proposed for and
were awarded phase curve observations
\citep[GO-14290,][]{crossfield:2019ddt} in both {\em Spitzer} channels
in  Cycle 14 DDT Review 2; these phase curve data are the
primary focus of this paper.

Our {\em Spitzer} observations provided near-continuous coverage of
one full orbital period of \planet\ in each of the 3.6\,\micron\ and
4.5\,\micron\ channels of the IRAC instrument \citep{fazio:2005}. Both
observations used standard best-practices for precise {\em Spitzer}
time-series photometry, with an initial peak-up observation designed
to place \starname\ on the well-characterized ``sweet spot'' of the
IRAC detector and observations bracketed by eclipses of the planet,
offerring an empirical means of calibrating out low-order detector and
stellar variability {\update (though ultimately we do not use these
  extra data)}.  Because of the relatively long duration of the
observations, each phase curve was split into two observing blocks
with a slight gap between them. Due to a scheduling oversight, in the
4.5\,\micron\ data set the gap partially overlaps with the transit of
\planet.

Because of the bright target star both channels of {\em Spitzer}
photometry used subarray-mode observations, which consist of multiple
sets of 64 quick subarray frames.  The 4.5\,\micron\ observations used
2\,s subarray integrations, with 303 subarray sets before the break
and 323 after. The 3.6\,\micron\ observations used 0.4\,s integrations,
with 1431 and then 1521 subarray sets. All raw data products were
automatically processed with version 19.2.0 of the {\em Spitzer} data
calibration pipeline before further analysis, and these data products
are publicly available through the Spitzer Heritage
Archive\footnote{\url{https://sha.ipac.caltech.edu/applications/Spitzer/SHA/}}.

  
\section{Light Curve Analysis}
\label{sec:lcanalysis}
To look for the infrared and optical phase curves of LTT~9779b, we
examined photometry of this system from both channels of Spitzer/IRAC
as well as from \TESS.  We extracted, calibrated, and modeled our {\em
  Spitzer} photometry using the Photometry for Orbits, Eclipses, and
Transits (POET\footnote{\url{https://github.com/kevin218/POET}}) code
\citep{stevenson:2012a,cubillos:2013}. To analyze the \TESS\ data we followed the
same approach described by \cite{daylan:2020}.

\subsection{Spitzer photometry, calibration, and model selection}
\begin{figure}[b!]
\includegraphics[width = 0.92\textwidth]{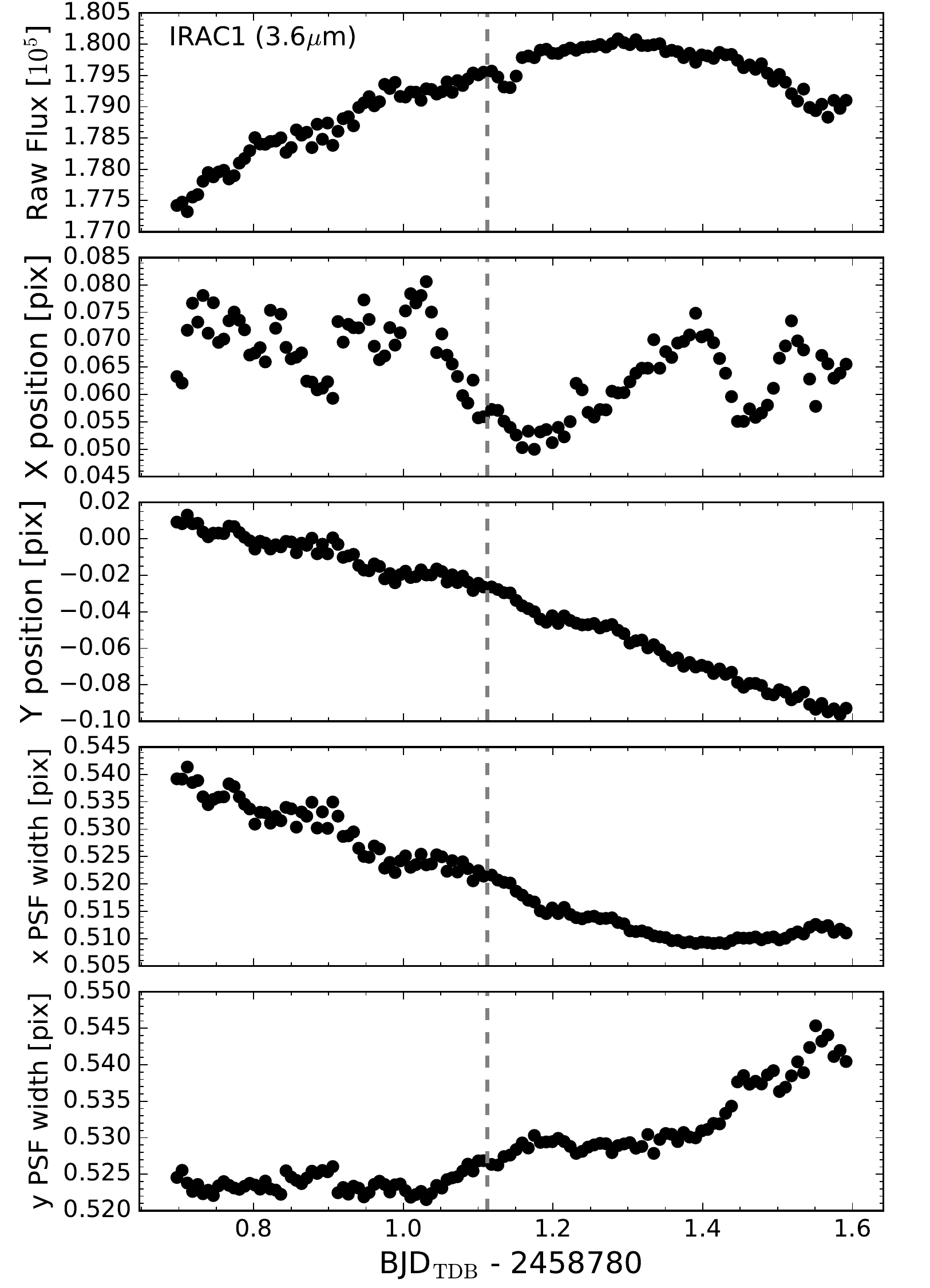}
\caption{From top to bottom: Spitzer/IRAC1 (3.6\,\micron) photometry,
  stellar $x$ and $y$ positions {\update relative to the (15, 15)
    pixel center}, and PSF $x$ and $y$ widths, for our observations of
  LTT~9779. The vertical dashed line indicates the break between the
  two AORs.  \label{fig:xy1}}
\end{figure}

\begin{figure}[b!]
\includegraphics[width = 0.92\textwidth]{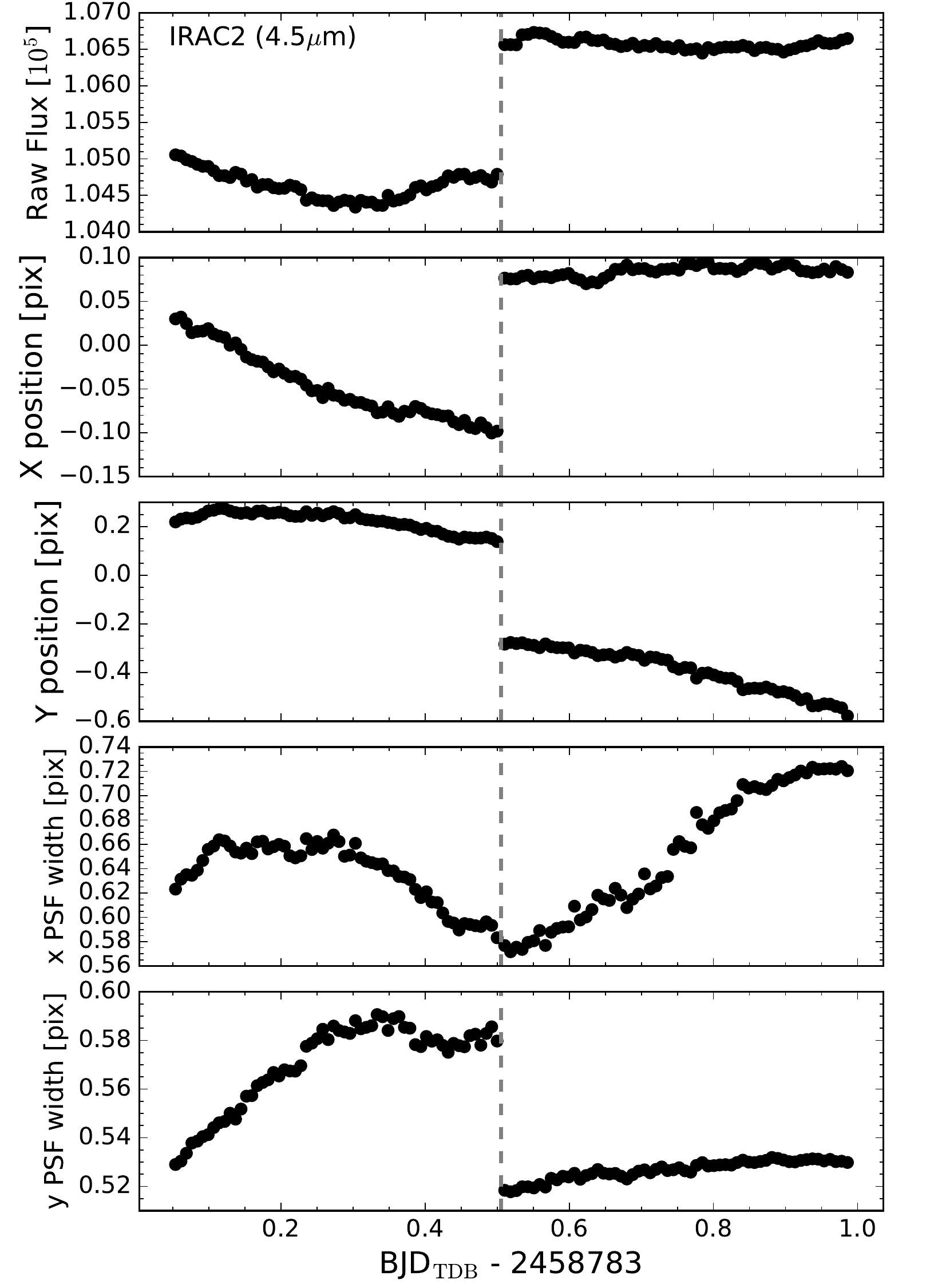}
\caption{From top to bottom: Spitzer/IRAC2 (4.5\,\micron) photometry,
  stellar $x$ and $y$ positions {\update relative to the (15, 15)
    pixel center}, and PSF $x$ and $y$ widths, for our observations of
  LTT~9779. The vertical dashed line indicates the break between the
  two AORs. \label{fig:xy2}}
\end{figure}

For the Spitzer data, we first used POET to calculate aperture
photometry for both IRAC channels with a range of aperture sizes, from
2.0 to 6.0~IRAC pixels and with inner and outer sky annulus radii of 7
and 15 pixels. Interpolated partial-pixel aperture photometry (with an
oversampling factor of five) accounted for fractional pixels in the
photometric calculations. To precisely track the location of the star
we fit a Gaussian profile to the stellar profile in each frame, using
a constant term to model each frame's background flux. The raw flux,
position, and PSF width are plotted in Figs.~\ref{fig:xy1}
and~\ref{fig:xy2} for the 3.6\,\micron\ and 4.5\,\micron\ data,
respectively.

POET models and removes IRAC's well-known systematic correlation
between a target's measured flux and the position of the source within
a pixel.  Without properly accounting for this intra-pixel sensitivity
effect, weak exoplanetary signals are (at best) difficult to
accurately extract.  POET accounts for this systematic effect using
its bilinearly-interpolated sub-pixel sensitivity (BLISS) mode
\citep{stevenson:2012a}. In BLISS mapping the main model
hyperparameters are the choice of grid size for the sub-pixel
sensitivity map, the astrophysical light curve model (e.g.\ phase
curve, eclipses, transits), and any additional light-curve models to
account for additional systematics (e.g.\ exponential ramp,
correlation with PSF width, or other long-term trends), as well as
the choice of photometric aperture size.

We followed past POET analyses of IRAC data by choosing the
hyperparameters that minimize the standard deviation of the normalized
residuals (SDNR) --- essentially minimizing the scatter on the
residuals to the full light-curve fit. However, to avoid overfitting
we also followed POET's recommendation to select the set of
hyperparameters that minimizes SDNR but for which a nearest-neighbor
interpolation does not outperform the BLISS interpolation.  After
testing a range of models, aperture sizes, and grid sizes (from 0.003
to 0.03 pixels), our final analysis identified the optimal set of
parameters for the 4.5\,\micron\ channel as a 5-pixel photometric
aperture, a 0.017~pixel grid scale, an astrophysical model including a
transit, secondary eclipse, and sinusoidal phase curve, and a
systematic model consisting of a quadratic correlation with PSF
width. As described below, this model does a good job at accounting
for the instrumental and astrophysical signals in the IRAC2
photometry; {\update  we achieve consistent final results with
  analyses using a four- or six-pixel apertures.}. In
Fig.~\ref{fig:phase2} we show the calibrated IRAC2 photometry and
best-fit light curve model. Fig.~\ref{fig:bindown2} shows that the
residuals to our fit bin down approximately as white noise. Based on
the WISE infrared flux of the star \citep{cutri:2014} and the IRAC
system throughput \citep{SSC:2012}, we calculate that our
4.5\,\micron\ photometry reaches 1.3$\times$ the expected photon noise
limit.

However, for the 3.6\,\micron\ IRAC1 photometry our models that
explain this channel's instrumental and astrophysical variations imply
a thermal phase curve with an implausibly large amplitude. These
optimal models include a 2--3~pixel aperture, a BLISS grid size of
0.005~pixels, a linear ramp at the start of the first AOR, a
correlation with PRF parameters, and a transit, secondary eclipse, and
sinusoidal phase curve.  In Fig.~\ref{fig:phase1} we show the
calibrated IRAC1 photometry and best-fit light curve model. Even after
calibration the light curve shows high scatter, little or no sign of
{\update a secondary eclipses at the start of the observation
  \citep[Eclipse 5 in the analysis of ][]{dragomir:2020}}, and the
best-fit phase curve model would nominally hint at negative planetary
flux between the transit and first eclipse.  Fig.~\ref{fig:bindown1}
shows that even when allowing nonphysical phase curve models, the
residuals to our light curve fits bin down substantially slower than
would be expected in the presence of white noise alone, further
indicating the presence of systematic correlated noise not captured by
the POET analysis.  We calculate that our systematics-corrected
3.6\,\micron\ photometry still reaches only 2.3$\times$ the expected
photon noise limit on short timescales, and substantially worse on
longer timescales.  Although the IRAC1 transit is recovered at high
confidence, due to this data set's intractable problems and high noise
levels we do not further consider the 3.6\,\micron\ eclipses
\citep[which is discussed in more detail by ][]{dragomir:2020} or
phase curve.





\subsection{Spitzer eclipse, transit, and phase curve modeling}
Although the main goal of this analysis is \planet 's phase curve
(which we detect at 4.5\,\micron), we also observed two transits (one
in each IRAC band).  The eclipse analysis is described in more detail
by \cite{dragomir:2020}, who analyzed our eclipse observations
together with four additional eclipses in each IRAC channel. Their
measurement of the eclipse depth is necessarily more precise than ours
would be, due to the larger number of eclipses analyzed. In our
analysis we therefore hold the eclipse depth fixed at their derived
value, \dayflux~ppm {\update (their analysis shows that the two
  eclipses shown in Fig.~\ref{fig:phase2} have a depth consistent at
  1.1$\sigma$ with the overall depth measured from all six
  eclipses)}. For other physical parameters (except where noted
below), we use the system parameters of \cite{jenkins:2020}.

Based on the stellar parameters from \cite{jenkins:2020}, we used the
3.6\,\micron\ and 4.5\,\micron\ quadratic limb-darkening coefficients from
\cite{claret:2013}.  The average limb-darkening coefficients
corresponding to $T_\mathrm{eff}=5500$~K and $\log g=4.5$ are
$u_1=0.106$ and $u_2=0.121$ (at 3.6\,\micron) and $u_1=-0.02$ and
$u_2=0.33$ (at 4.5\,\micron). As noted previously, our IRAC2 primary transit
occurs during a break in our Spitzer observations. This break makes
the transit somewhat more challenging to model, so we elect to keep
$u_2$ fixed at the tabulated value and allow $u_1$ to float within the
range of values calculated by \citep{claret:2013}, i.e.\ in the range
from $-0.13$ to $+0.09$.

For our phase curve modeling, we use a simple sinusoidal model
with functional form
\begin{equation}
  f(\phi) = 1 - A \cos (\phi - \Delta\phi)
  \label{eq:sinusoid}
\end{equation}
where $\phi$ is the orbital phase (zero at transit, $\sim$0.5 at
secondary eclipse), $A$ is the normalized (fractional) semi-amplitude,
and $\Delta \phi$ is the phase offset (i.e., the planetary longitude
corresponding to maximum flux). Since our final S/N on \planet 's
IRAC2 phase curve is relatively low, more complicated models (e.g. including
higher-order harmonics) were not justified by the data.


\subsection{TESS Phase Curve Analysis}
\label{sec:tessphase}

We also modeled the PDC (Pre-Search Data Conditioning) light curve of
LTT 9779b \citep{smith:2012pdc,stumpe:2012,stumpe:2014} using the
methodology presented by \cite{daylan:2020} to infer the planet's
phase modulation components at optical wavelengths.

We performed a global fit of the full PDC light curve using
\texttt{allesfitter} \citep{guenther:2019,guenther:2020}, an inference
framework for joint modeling of light curve and radial velocity
data. We modeled the light curve as the sum of the stellar baseline
emission occulted by the planet during the primary transit, the
planetary baseline emission occulted by the star during the secondary
eclipse, the planetary modulation in the form of a cosine at the
orbital period that peaks at superior conjunction, the ellipsoidal
variation in the form of a cosine at half the orbital period that
peaks at quadrature phases, and an additive constant offset to absorb
any normalization bias. Although the planetary modulation can be
subject to a phase shift, we nevertheless assumed a phase shift of
zero (the value expected for reflected light and for thermal emission
from such a hot planet). This is because the resulting planetary
modulation amplitude was small, leaving the optical phase shift
unconstrained. We also assumed a circular orbit.  We modeled the data
(which could consist of thermal emission, reflected starlight, or
both) as a linear combination of the following components:

\begin{itemize}
\item planetary day side: $\frac{A_d}{2} (1 - \cos 2\pi\phi)$ outside the secondary eclipse and 0 otherwise,
\item planetary night side:, $\frac{A_n}{2} (1 + \cos 2\pi\phi)$ outside the secondary eclipse,
\item ellipsoidal variation, $A_e \sin 4\pi \phi$, i.e., a sinusoid at twice the orbital period that peaks at quadrature (0.25 and 0.75) phases,
\item an additive constant offset to absorb any normalization bias.
\end{itemize}
$A_d$, $A_n$ and $A_e$ parameterize the amplitudes of these
components.

We sampled from the posterior distribution of the global model
parameters using uniform priors. The posterior median phase curve
components are shown in Fig.~\ref{fig:tessphase}, where the black
points denote the light curve phase-folded at the posterior median
period and binned. The blue line shows the posterior median model
fitted to the data. The colored dashed lines show the individual
components of the posterior median model. The stellar baseline,
planetary baseline, planetary modulation, and the ellipsoidal
variation are shown in Figure 1 with orange, olive, magenta, and red
colors, respectively.  We obtain a secondary eclipse depth of
$59^{+24}_{-21}$~ppm \citep{dragomir:2020}, a phase curve amplitude of
$25^{+14}_{-13}$~ppm, and a night-side emission of
$33^{+24}_{-20}$~ppm. The photometric precision we achieve is in line
with expectations from the demonstrated performance of \TESS: The star
is \TESS\ mag$=9$, so the expected photometric precision is roughly
100~ppm hr$^{-1/2}$ \citep{ricker:2014}.  The phase curve plotted in
Fig.~\ref{fig:tessphase} has been averaged down into 100 bins, each
containing roughly 150 2-minute cadence measurements -- i.e., 5~hr. So
the expected 1$\sigma$ precision is $100/\sqrt(5) \approx 45$~ppm,
consistent with our achieved precision. {\update
  Fig.~\ref{fig:bindowntess} shows how the residuals to our {\em TESS}
  analysis bin down over time, which suggests residual correlated
  noise of no more than 15\% on the timescale of LTT~9779b's transit
  duration.}


\section{Measurement and Interpretation}
\label{sec:measure}

\subsection{Spitzer Transits and an Updated Ephemeris}
Our IRAC transits are mainly useful for refining the planet's orbital
ephemeris and reducing the uncertainty on its orbital period from that
reported in the initial \TESS\ analysis of \cite{jenkins:2020} and the
updated analysis of \cite{dragomir:2020}. Our 3.6\,\micron\ and
4.5\,\micron\ transit analyses yield mid-center times of
$T_0=\transittimeone \pm \utransittimeone$ and $\transittimetwo \pm
\utransittimetwo$, respectively (BJD$_\mathrm{TDB}$). The IRAC1 and
IRAC2 transits occurred 539 and 542 orbits, respectively, after the
discovery epoch \citep{jenkins:2020}, which reported
$T_0=2458354.21430 \pm 0.00025$ and $P=0.7920520 \pm 0.0000093$. With
our new Spitzer transit times, a weighted least-squares fit gives an
updated mid-transit time of $\transittime \pm \utransittime$ {\update
  (BJD$_\mathrm{TDB}$)} and a refined period of $\newperiod \pm
\unewperiod$\,d, a measurement 13.6$\times$ more precise than the
discovery period.

Our 3.6\,\micron\ and 4.5\,\micron\ transit depths are
$R_p/R_*=\rprsone \pm \urprsone$ and\ $\rprstwo \pm \urprstwo$,
respectively, consistent with the {\em TESS} transit depth
\citep{jenkins:2020}.  Neither the infrared nor the optical transit
measurements are sufficiently precise to usefully constrain \planet 's
atmosphere in transmission, because the planet's expected scale height
is relatively small \citep{jenkins:2020}.


\hspace{-1in}
\begin{deluxetable}{l l l l l }[bt]
\hspace{-1in}\tabletypesize{\scriptsize}
\tablecaption{  Properties of \planet \label{tab:params}}
\tablewidth{0pt}
\tablehead{
  &&&&\\
  \colhead{Parameter} & \colhead{Units}    & \colhead{Value} & \colhead{1$\sigma$ Uncertainty} &  \colhead{Notes} \\
}
\startdata
\multicolumn{5}{l}{\hspace{0.2in}\em Fit Parameters:}\\
$R_p/R_*$ & --- & \rprsone& $\pm$\urprsone & (3.6\,\micron; this work) \\
$R_p/R_*$ & --- & \rprstwo& $\pm$\urprstwo & (4.5\,\micron; this work) \\
$T_0$  & BJD$_\mathrm{TDB}$ & \transittimeone & $\pm$\utransittimeone & (3.6\,\micron; this work) \\ 
$T_0$  & BJD$_\mathrm{TDB}$ & \transittimetwo & $\pm$\utransittimetwo & (4.5\,\micron; this work) \\ 
$A$ & ppm & \phaseamplitude & $\pm$\uphaseamplitude & (4.5\,\micron; this work)  \\
$\Delta\phi$\tablenotemark{a} & deg & \phaseoffset& $\pm$\uphaseoffset& (4.5\,\micron; this work) \\ \hline
\multicolumn{5}{l}{\hspace{0.2in}\em Derived Parameters:} \\
$T_0$ & BJD$_\mathrm{TDB}$ & \transittime & $\pm$\utransittime & This work\\
$P$ & d & \newperiod& $\pm$\unewperiod& This work\\
$T_\mathrm{day}$ & K & \thot& $\pm$\uthot & \citep[4.5\,\micron, from][]{dragomir:2020}\\
$T_\mathrm{night}$ & K & \tcold & $\pm$\utcold & (4.5\,\micron; this work) \\
$\Delta T$ & K &\deltat & $\pm$\udeltat & (4.5\,\micron; this work)\\ 
\enddata
\tablenotetext{a}{{\update This} best-fit phase offset is slightly west of the substellar point.}
\end{deluxetable}

\subsection{Phase curves}

Our phase curve model (Eq.~\ref{eq:sinusoid}) can be used to determine
the hemisphere-averaged flux and brightness temperature from the
planet's day and night sides. To effect this conversion, we use the
BT-Settl library of synthetic stellar spectra \citep{allard:2014} to
properly model the non-blackbody stellar flux. We perform this
calculation in a Monte Carlo framework in order to properly
account for the uncertainties in both the secondary eclipse and phase
curve amplitude. As expected, we also find that the large
uncertainties on the phase curve parameters dominate the much smaller
uncertainties in the stellar parameters.

We calculate 4.5\,\micron\ day and night-side temperatures of
$T_\mathrm{day}=\thot\pm\uthot$ and
$T_\mathrm{night}=\tcold\pm\utcold$, with upper limits on the
night-side brightness temperature of $<1350$~K and $<1650$~K at 95.4\%
and 99.73\% confidence (2$\sigma$ and 3$\sigma$), respectively. Note
that our measurement of the day-side brightness temperature is
somewhat more precise than that derived from the secondary eclipse
analysis of \citep{dragomir:2020}, since our analysis also includes
the full phase curve.

Combining the posterior distributions of the day- and night-side
temperatures, we find a day-night temperature contrast of $\Delta
T=\deltat\pm\udeltat$~K.  The distribution of $\Delta T$ is symmetric
but non-Gaussian, with 95.4\% and 99.73\% (2$\sigma$ and 3$\sigma$)
confidence intervals of $\pm\utwodeltat$~K and $\pm\uthreedeltat$~K,
respectively.  We also measure the planet's phase offset (the
longitude of maximum flux) to be $\phaseoffset^{\rm
  o}\pm\uphaseoffset^{\rm o}$ {\update east} of the substellar point.  Our full
set of phase curve measurements from our {\em Spitzer} analysis are
listed in Table~\ref{tab:params}.

In our \TESS\ light curve analysis, we sampled from the posterior
distribution of $A_d$, $A_n$ and $A_e$ (see Sec.~\ref{sec:tessphase})
using uniform priors. The posterior median of these components are
plotted against the \TESS\ photometry in Fig.~\ref{fig:tessphase},
along with the posterior median of the total model.  As described by
\cite{dragomir:2020}, we find a TESS secondary eclipse depth of
$55^{+24}_{-21}$ ppm (a roughly 3-$\sigma$ detection). However, the
planet's phase variation is not detected at high significance: we find
a phase amplitude in the TESS bandpass of just $29\pm17$ ppm. Further
\TESS\ observations of the system, anticipated in the \TESS\ extended
mission, would naturally improve the precision of these measurements.

\subsection{General Circulation Models}


\hspace{-1in}
\begin{deluxetable}{l l | c c | c c}[bt]
\hspace{-1in}\tabletypesize{\scriptsize}
\tablecaption{  Phase Curve Properties \label{tab:gcmcomparison}}
\tablewidth{0pt}
\tablehead{
  &&&&\multicolumn{2}{|c}{}\\
  \colhead{Bandpass} & \colhead{Source}    & \multicolumn{1}{|c}{Amplitude\tablenotemark{a}} & \colhead{Offset\tablenotemark{b}} & \multicolumn{2}{|c}{$F_P/F_*$ [ppm] at:}  \\
\colhead{} & \colhead{} & \multicolumn{1}{|c}{[ppm]} & \colhead{[deg]} & \multicolumn{1}{|l}{$\phi=0$} & \colhead{$\phi=0.5$} }  

\startdata
TESS & data               & $29 \pm 17$ & ---\tablenotemark{c} &$33^{+24}_{-20}$ & $59_{-21}^{+24}$ \tablenotemark{d} \\
     & Solar abundances   & 1.0 & 77  & 2.5 & 2.6 \\
     & $30\times$ Solar   & 7.8 & 41  & 4.7 & 10.7 \\
Spitzer 4.5\micron & data & $\phaseamplitude \pm \uphaseamplitude$  & $\eastphaseoffset \pm \uphaseoffset$  &$\nightflux \pm \unightflux$ & $\dayflux\pm \udayflux$ \tablenotemark{d} \\
     & Solar abundances   & 113 & 60  & 241 & 291 \\
     & $30\times$ Solar   & 261 & 23  & 272 & 511 \\
\enddata
\tablenotetext{a}{Peak-to-valley phase curve amplitude.}
\tablenotetext{b}{Eastward shift of the phase curve maximum from the substellar point.}
\tablenotetext{c}{Zero phase offset assumed; see Sec.~\ref{sec:tessphase}.}
\tablenotetext{d}{From \cite{dragomir:2020}.}
\end{deluxetable}







To better interpret our observations of LTT~9779b, we calculated 3D
general circulation models (GCMs) of this planet utilizing the
SPARC/MITgcm \citep[e.g.,][]{showman:2009}.  To reflect the range of
possible observed atmospheric compositions we used two different model
configurations, with atmospheric elemental abundances at the Solar
level and 30$\times$ the solar level.  Note that our simulations do
not include the effect of atmospheric condensates, which some studies
suggest could be important for planets even at these high temperatures
\citep[e.g.,][]{parmentier:2016a}.  In particular, atmospheric
aerosols would likely increase the geometric albedo in the
\TESS\ bandpass and so bring our GCM predictions into better agreement
with the data at these shorter wavelengths. 

Fig.~\ref{fig:gcmcomparison} compares the 4.5$\mu$m and \TESS\ phase
curves predicted by our GCMs with our observations, and some relevant
values for comparison are also enumerated in
Table~\ref{tab:gcmcomparison}. At optical wavelengths, our GCMs
predict lower flux than we observe with \TESS\ at all orbital
phases. At 4.5\,\micron, when compared to the observations the
30$\times$ Solar GCM predicts a hotter day side and smaller phase
offset, while the Solar-abundance GCM predicts a somewhat cooler day
side, smaller day-night temperature contrast, and larger phase
offset. The trend seen in Fig.~\ref{fig:gcmcomparison} of a phase
curve amplitude that increases with atmospheric metallicity mirrors
that seen in similar simulations for the similarly-sized, but much
cooler, warm Neptune GJ~436b \citep{lewis:2010}.  Nonetheless, neither
of our GCMs provides an especially good match to our observations,
which exhibit a much colder night side and smaller (consistent with
zero) phase offset. Indeed, {\update if anything the phase curve from
  the Solar-abundance GCM is a better match to our observations than
  is the 30$\times$ Solar model.  And} although our constraint on the
phase offset is fairly loose, our measured day-side flux and phase
curve amplitude both disagree with those predicted by the GCMs.

In Fig.~\ref{fig:phasecontrast}, we also compare our measurements of
the planet's day- and night-side emission to the phase-resolved
emergent spectra predicted by the GCMs.  These reinforce the
impressions conveyed by the comparison in Fig.~\ref{fig:gcmcomparison}
and Table~\ref{tab:gcmcomparison}: at 4.5\,\micron\ the
Solar-abundance GCM over-predicts the night-side flux but
under-predicts the day-side flux, while the 30$\times$-solar GCM
over-predicts the planet's flux at all phases; {\update our
  (aerosol-free)} GCMs underestimate the flux emitted in the TESS
bandpass {\update by 2.1$\sigma$ and 2.4$\sigma$ (for the Solar and
  30$\times$ Solar abundance GCMs, respectively)}. On balance, the
30$\times$-solar GCM matches our observations somewhat better {\update
  though (unlike Fig.~\ref{fig:gcmcomparison}) the comparison in
  Fig.~\ref{fig:phasecontrast} includes the 3.6\,\micron\ day-side
  (eclipse) measurement and so more strongly favors the
  30$\times$-solar GCM ($\Delta \chi^2 = 15$), hinting at a
  super-Solar metallicity for LTT~9779b's atmosphere.} However,
Fig.~\ref{fig:phasecontrast} also demonstrates that this model (unlike
the Solar-abundance model) predicts a nearly isothermal day-side
photosphere with only very weak spectral features, contrary to the
strong absorption feature implied by the discrepant planetary
brightness temperatures at 3.6\,\micron\ and 4.5\,\micron.

\section{Discussion and Interpretation}
\label{sec:discinterp}
\subsection{Phase Curve interpretation}
The large phase-curve amplitude and small phase offset in our
4.5\,\micron\ phase curve both suggest that LTT~9779b has a
high-metallicity atmosphere.  Higher-metallicity atmospheres have
higher opacities, and Neptune-size planets may have atmospheric
metallicities of 100$\times$ Solar or more
\citep{fortney:2013,wakeford:2017}, much greater than expected for hot
Jupiters. Therefore, the atmospheric layers probed by thermal emission
measurements may tend to be at lower pressures for Neptunes than for
Jovians, with consequently shorter radiative timescales. As a result,
the phase curve offsets are expected to be smaller and the phase
amplitudes larger than for the lower-metallicity hot Jupiters
\citep{parmentier:2018}. Our phase offset measurement is not precise
enough for a useful comparison, but the enhanced phase amplitude that
we see suggests that LTT~9779b's atmosphere may have an enhanced
metallicity. The host star has a super-solar metallicity of
[Fe/H]=+0.25~dex, but based on the GCM simulations shown in
Fig.~\ref{fig:gcmcomparison} the planet's metallicity would be much
more metal-rich, with [Fe/H]$\gtrsim1.5$~dex.

We also used these brightness temperature with the radiative-balance
model of \cite{cowan:2011} to estimate LTT~9779b's Bond albedo $A_B$
and global heat recirculation efficiency $\epsilon$, as shown in
Fig.~\ref{fig:albedoposteriors}a.  In this framework, $\epsilon$
equals zero or unity, respectively, for zero circulation or full
day-night recirculation.  We find
$\epsilon$=\recirc$\pm$\urecirc\ (2$\sigma$ upper limit of
$<\uppertworecirc$), indicating a very low level of heat
redistribution consistent with the planet's near-zero phase offset.
For LTT~9779b's Bond albedo we find a surprisingly high value of
$A_B$=\ab$\pm$\uab\ {\update from our 4.5\,\micron\ data}. Presumably
this $A_B$ value is so high because {\update CO and/or CO$_2$ absorb
  strongly in this planet's atmosphere}
4.5\,\micron\ \citep{dragomir:2020}, and so relatively less emission
is seen in our IRAC2 observations. In this case, the high $A_B$ we
derive would reflect the limits of single-band photometry for detailed
energy balance calculations. \cite{cowan:2011} found that a single
broadband infrared brightness temperature measurement translates into
a roughly 8\% systematic uncertainty in the effective temperature, and
hence into a roughly 32\% systematic uncertainty in bolometric flux
and the Bond albedo.  {\update As a check on
  $A_B$, we repeated the calculation using the 3.6\,\micron\ eclipse
  measurement and assuming all 3.6\,\micron\ flux is emitted on the
  day side.  In this case we calculate $A_B= 0.27\pm 0.16$ (or
  $0.27^{+0.41}_{-0.26}$ at 99.73\% confidence), which would not be
  remarkable in the context of the many hot Jupiters that have been
  studied in this way (see Fig.~\ref{fig:phasecurvesample}f).}


On the other hand, LTT~9779b has $(R_P/a)^2=142$~ppm, which together
with the \TESS\ secondary eclipse indicates a geometric albedo in the
\TESS\ bandpass of roughly 0.4 if all the optical light were the
result of scattering. A 2100~K blackbody would contribute only roughly
10~ppm of thermal emission at these wavelengths (see
Fig.~\ref{fig:gcmcomparison}b), though such a planet's spectrum is
unlikely to closely resemble a blackbody. The optical data may
therefore hint at a moderately high {\update geometric albedo, which
  would differ qualitatively from the near-zero geometric albedo
  measured for some hot Jupiters
  \citep[e.g.\ WASP-12b;][]{bell:2017}}. Nonetheless, better flux
measurements are needed to confirm this point.

Finally, we used a similar modeling approach to evaluate LTT~9779b's
4.5\,\micron\ phase curve in terms of an alternative two-parameter
model in which a planet's global temperature distribution depends on
$A_B$ and the ratio of radiative to advective timescales,
$\tau_\mathrm{rad}/\tau_\mathrm{adv}$ \cite[][; see the Appendix for some
  pedagogically useful analytic insights associated with this model]{cowan:2011circ}.  The posterior distribution of
these parameters is shown in Fig.~\ref{fig:albedoposteriors}b,
indicating $A_B$=\abtwo$\pm$\uabtwo\ (consistent with the measurement
described immediately above) and a 2$\sigma$ upper limit of
$\tau_\mathrm{rad}/\tau_\mathrm{adv} < \radadvupperlimit$ (consistent
with expectations for such a hot planet).

To put LTT~9779b's phase curve and derived parameters in the context
of other 4.5\,\micron\ Spitzer phase curves, we followed the same
procedures as described above to calculate $T_\mathrm{day}$,
$T_\mathrm{night}$, $\Delta T$, $A_B$, and $\epsilon$ for all planets
with published IRAC2/4.5\,\micron\ phase curves. These measurements
are shown in Fig.~\ref{fig:phasecurvesample}, along with the
predictions for LTT~9779b from our GCMs.  We restrict this comparison
to the sixteen planets with nearly-circular orbits
\citep{knutson:2012,maxted:2012,zellem:2014,demory:2016b,wong:2015,wong:2016,zhang:2018,
  kreidberg:2018,kreidberg:2019,bell:2019,stevenson:2017,dang:2018,mansfield:2020,keating:2020}.

Fig.~\ref{fig:phasecurvesample} allows us to examine the sample of
4.5\micron\ exoplanetary phase curves for possible trends between
these planetary parameters.  For example, \cite{keating:2019} and
\cite{beatty:2019} claimed that the roughly uniform, $\sim$1100~K
night-side temperatures of a sample of twelve hot Jupiters indicated
the ubiquitous presence of night-side clouds. Our updated sample shows
that this trend generally continues to hold, with the exceptions of
hot Jupiters CoRoT-2b and KELT-9b
\citep{dang:2018,mansfield:2020} and much smaller,
presumed-rocky LHS~3844b \citep[which has at most a tenuous
  atmosphere][]{kreidberg:2019}.  In a similar vein, \cite{zhang:2018}
used Spitzer phase curves of ten hot Jupiters to claim a coherent
trend between $T_\mathrm{irr}$ and eastward phase offsets, which would
indicate the increasing importance of magnetohydrodynamical effects at
higher temperatures as well as high-altitude clouds.  However, several
recent measurements do not seem to support this trend, in particular
the low phase offsets of Qatar-1b \citep[consistent with zero shift at
  2000~K][]{keating:2020} and CoRoT-2b \citep[a westward shift of
  $23\pm4^\mathrm{o}$ at 2100~K][]{dang:2018}.




With two exceptions, the ensemble of 4.5\,\micron\ phase curve
measurements depicted in Fig.~\ref{fig:phasecurvesample} show that the
parameters of LTT~9779b derived from its phase curve are generally
similar to those of the population of hot Jupiters.  The two
exceptions, which stem from a single explanation, are that LTT~9779b
exhibits an unusually low day-side temperature, and therefore has a
{\update high Bond albedo as derived from 4.5\,\micron\ observations
  (as discussed above)}, when compared to hot Jupiters with similar
irradiation temperatures. These discrepancies are due to the
4.5\,\micron\ absorption feature inferred in LTT~9779b's day-side
spectrum \citep{dragomir:2020}. In contrast, most hot Jupiters at
these temperatures exhibit more nearly blackbody-like day-side
emission spectra due to shallower photospheric temperature gradients.

Observations inside molecular bands (e.g. the CO/CO$_2$ band at 4.5
$\mu$m) probe shallower atmospheric layers. This implies that our 3.6
$\mu$m phase curve should show a larger phase offset relative to the
4.5 $\mu$m phase curve, since it probes deeper into the atmosphere
where energy re-radiation is less rapid and advective heat transport
is more efficient. It is tempting to interpret the large phase offset
apparently seen in the 3.6\,\micron\ phase curve
(Fig.~\ref{fig:phase1}) as evidence of this phenomenon, but the
nonphysical models required to explain this photometry (along with its
high levels of correlated noise) caution against such an
interpretation. Nevertheless, phase curves of this planet at
additional wavelengths, probing a broader range of pressures, would
provide a powerful diagnostic for measuring thermal structure and
chemical abundances across the entire planet.

A particularly interesting comparison is between LTT~9779b and
WASP-19b, an inflated hot Jupiter with multi-band eclipse and phase
curve measurements \citep[][and references therein]{wong:2016}.  The
surface gravity and irradiation temperatures of these two planets are
within 10\% of each other, as shown in Table~\ref{tab:hotplanets}.
Given two {\em stars} with such similar $T_\mathrm{eff}$ and $\log g$,
one would expect any spectral differences to result solely from
different metallicities; the same argument should presumably apply to
planets as well.  Yet WASP-19b has an essentially featureless day-side
spectrum (unlike LTT~9779b) and the brightness temperature differences
in the two Warm Spitzer bandpasses, $T_\mathrm{day}(3.6\,\micron) -
T_\mathrm{day}(4.5\,\micron)$, is greater for LTT~9779b than for
WASP-19b at 99.1\% confidence\footnote{During the review process
  another analysis of WASP-19b's thermal emission was released
  \citep{rajpurohit:2020} that reports the planet's atmosphere hosts a
  thermal inversion, further strengthening the difference between it
  and LTT~9779b.}.  Table~\ref{tab:hotplanets} also compares our hot
Neptune to WASP-14b, another similarly-irradiated hot Jupiter but with
10$\times$ greater surface gravity. The case here is similar, with the
two-color brightness temperature difference greater for LTT~9779b at
98.2\% confidence. {\update Although both additional data and further
  modeling are warranted, the different {\em Spitzer} emission spectra
  of LTT~9779b and WASP-14b \&\ WASP--19b may be explained by the
  tendency of higher-metallicity atmospheres to show more
  strongly-decreasing thermal profiles at $T_\mathrm{eq} \lesssim
  2300$~K \citep{mansfield:2020ep3}.  Thus this comparison with hot
  Jupiters provides another tentative} line of evidence that our
target may have a significantly different (presumably higher)
atmospheric metallicity than do hot Jupiters\footnote{During the
  review process another population study of {\em Spitzer}/IRAC
  eclipses of hot Jupiters was released \citep{baxter:2020}, reporting
  that planets with similar irradiation levels to LTT~9779b have
  brightness temperature ratios
  $T_\mathrm{day}(4.5\,\micron)/T_\mathrm{day}(3.6\,\micron) \gtrsim
  1$, while our data for LTT~9779b show a ratio of $0.78\pm0.07$ ---
  further evidence that this planet is unlike hot Jupiters.}.

\begin{deluxetable}{l c c c}[bt]
\hspace{-1in}\tabletypesize{\scriptsize}
\tablecaption{  Brightness Temperatures of Three Hot Exoplanets \label{tab:hotplanets}}
\tablewidth{0pt}
\tablehead{&&&\\
\colhead{Planet} &  \colhead{WASP-14b} & \colhead{WASP-19b} & \colhead{LTT~9779b}  \\
}
\startdata
Irradiation Temperature  (K)         & $2640 \pm 43$   & $2990 \pm 50$      & $2760 \pm 33$ \\
Planet Mass   ($M_\mathrm{Jup}$)      & $ 7.3\pm0.5 $    & $1.114 \pm 0.036$  & $0.0922 \pm 0.0025$ \\
Planet Radius ($R_\mathrm{Jup}$)      & $1.28 \pm0.08 $  & $1.395 \pm 0.023$  & $0.421 \pm 0.021 $ \\
Surface Gravity  ($\log_{10}$ [cgs])  & $4.107 \pm 0.043$& $3.152 \pm 0.020$ &  $3.110 \pm 0.044$ \\ \hline
 $T_\mathrm{day}$ (3.6\,\micron)   (K) & $2341 \pm 37$   & $2361 \pm 48$      & $2305 \pm 141$ \\  
 $T_\mathrm{day}$ (4.5\,\micron)   (K) & $2241 \pm 45$   & $2331\pm 67$       & $\thot \pm \uthot$ \\
 $T_\mathrm{night}$ (4.5\,\micron) (K) & $1301 \pm 69$   & $1130 \pm 320$     & $\tcold \pm \utcold$ \\
 $T_\mathrm{day}(3.6\,\micron) - T_\mathrm{day}(4.5\,\micron)$ (K) & $100 \pm 58$ & $30 \pm 82$    & $510 \pm 180$ \\ 
 $T_\mathrm{day} - T_\mathrm{night}$ (4.5\,\micron)            (K) & $940 \pm 78$ & $1200 \pm 330$ & $\deltat \pm \udeltat$
\enddata
\tablenotetext{a}{Measurements taken from \cite{wong:2015} (WASP-14b), \cite{wong:2016} (WASP-19b), and \cite{dragomir:2020} and this work (LTT~9779b).}
\end{deluxetable}

\subsection{Conclusions and Future Prospects}

One of the key goals of the ongoing TESS survey is to identify the
best exoplanet targets for detailed atmospheric characterization
\citep{ricker:2014}.  TESS is already discovering such systems,
opening wider the door to atmospheric studies of smaller planets.
Among the most exciting TESS planets are those for which existing
facilities can already answer new questions about the atmospheres of
new classes of objects or of individual planets. LTT~9779 is such a
system.

\begin{figure}[b!]
\includegraphics[width = 0.92\textwidth]{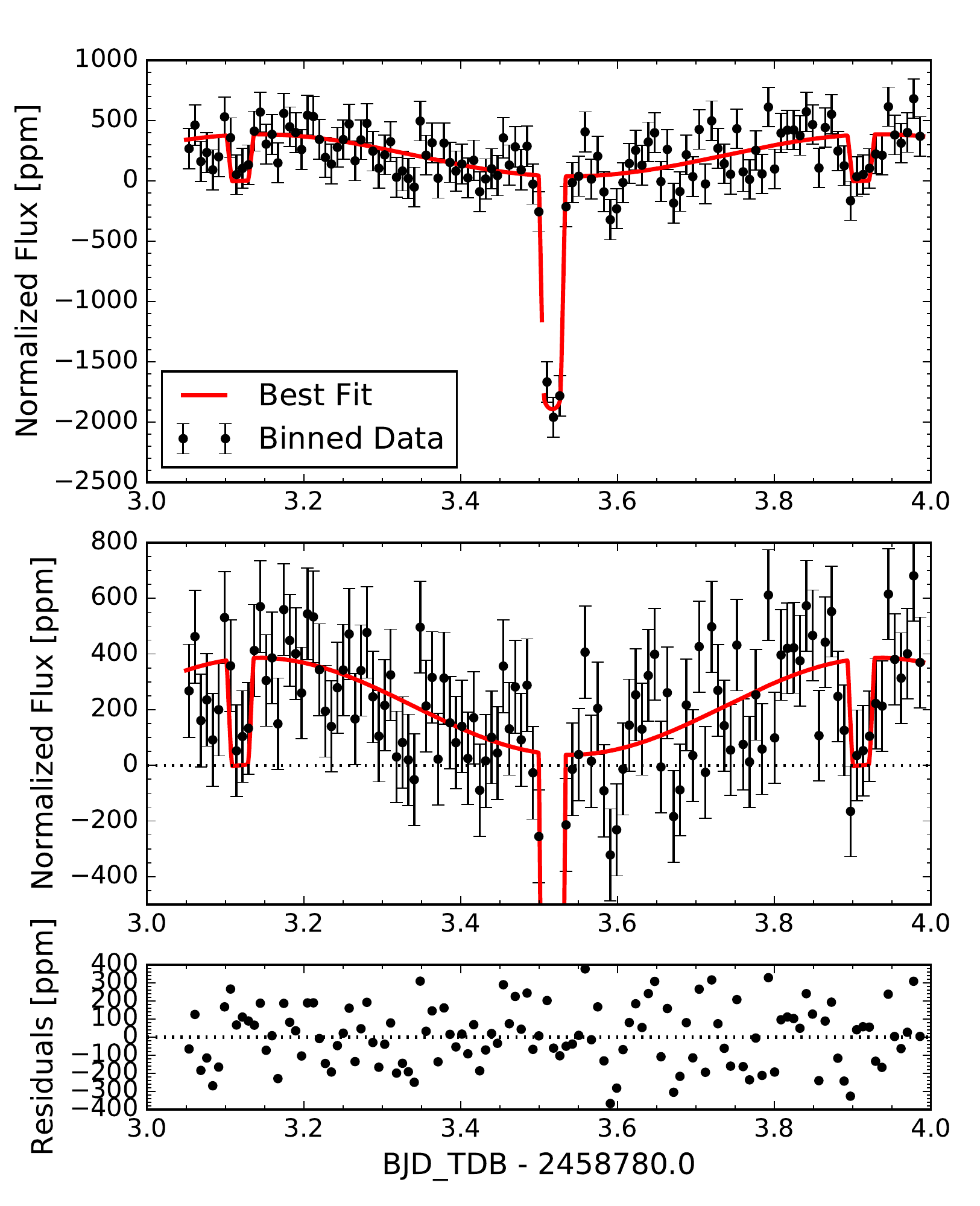}
\caption{Spitzer/IRAC2 (4.5\,\micron) observations and phase curve.
  From top to bottom: the full set of observations, the full data set
  again (but stretched to highlight the phase variations), and the
  residuals to the fit. The binned photometry is indicated by the
  black points, and the red line is the best-fit model. These data
  reach 1.3$\times$ the photon-noise limit on short timescales and
  exhibit no sign of residual correlated noise
  (Fig.~\ref{fig:bindown2}). \label{fig:phase2}}
\end{figure}

\begin{figure}[b!]
\includegraphics[width = 0.92\textwidth]{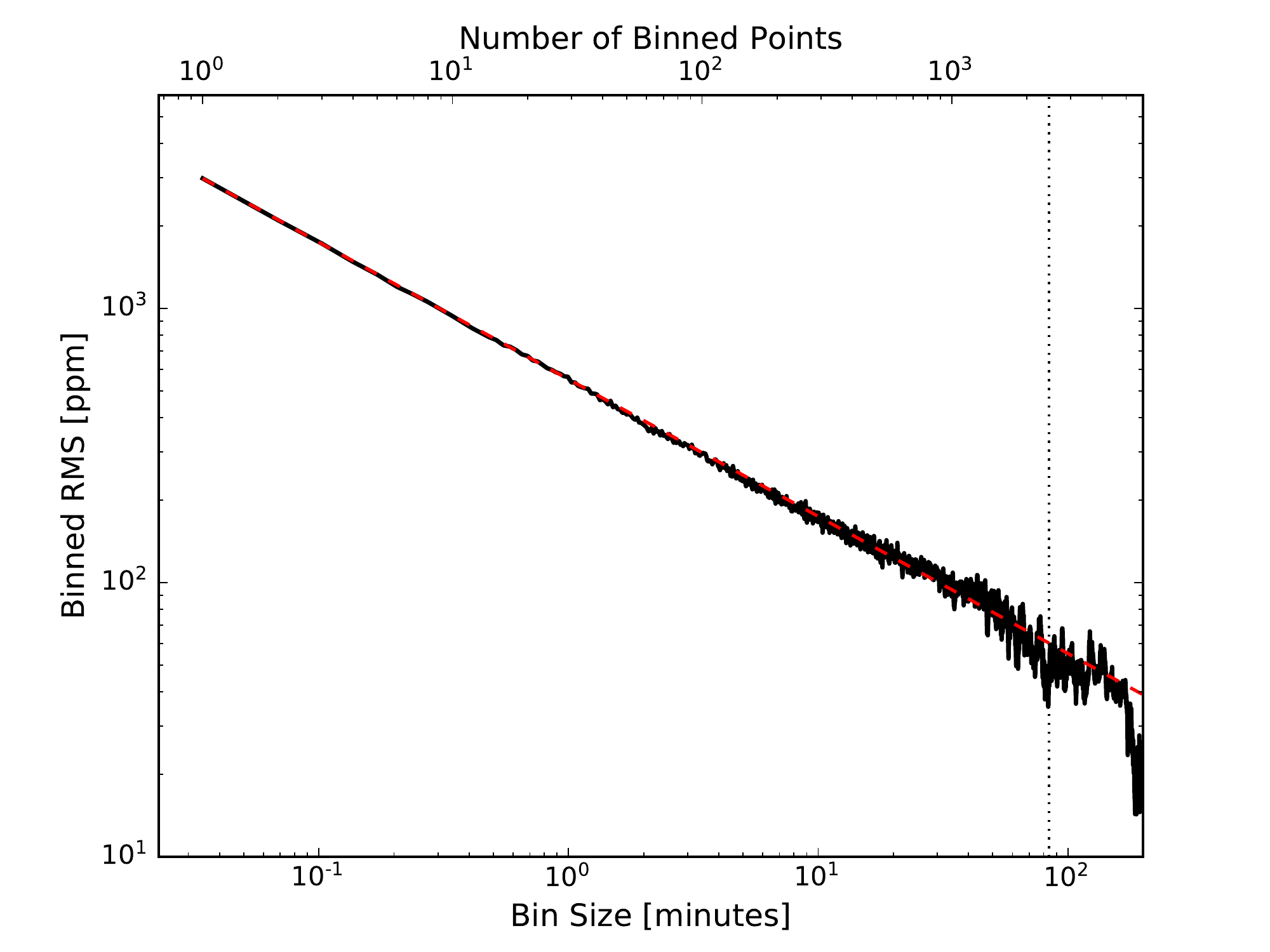}
\caption{The dispersion of the binned residuals (solid line) to the
  4.5\,\micron\ light curve (Fig.~\ref{fig:phase2}) shows no significant
  evidence for correlated noise. The dashed line shows the expectation
  for wholly uncorrelated errors, which scale as $N^{-1/2}$, and the
  vertical dotted line indicates the transit duration.
  \label{fig:bindown2}}
\end{figure}

\begin{figure}[b!]
\includegraphics[width = 0.92\textwidth]{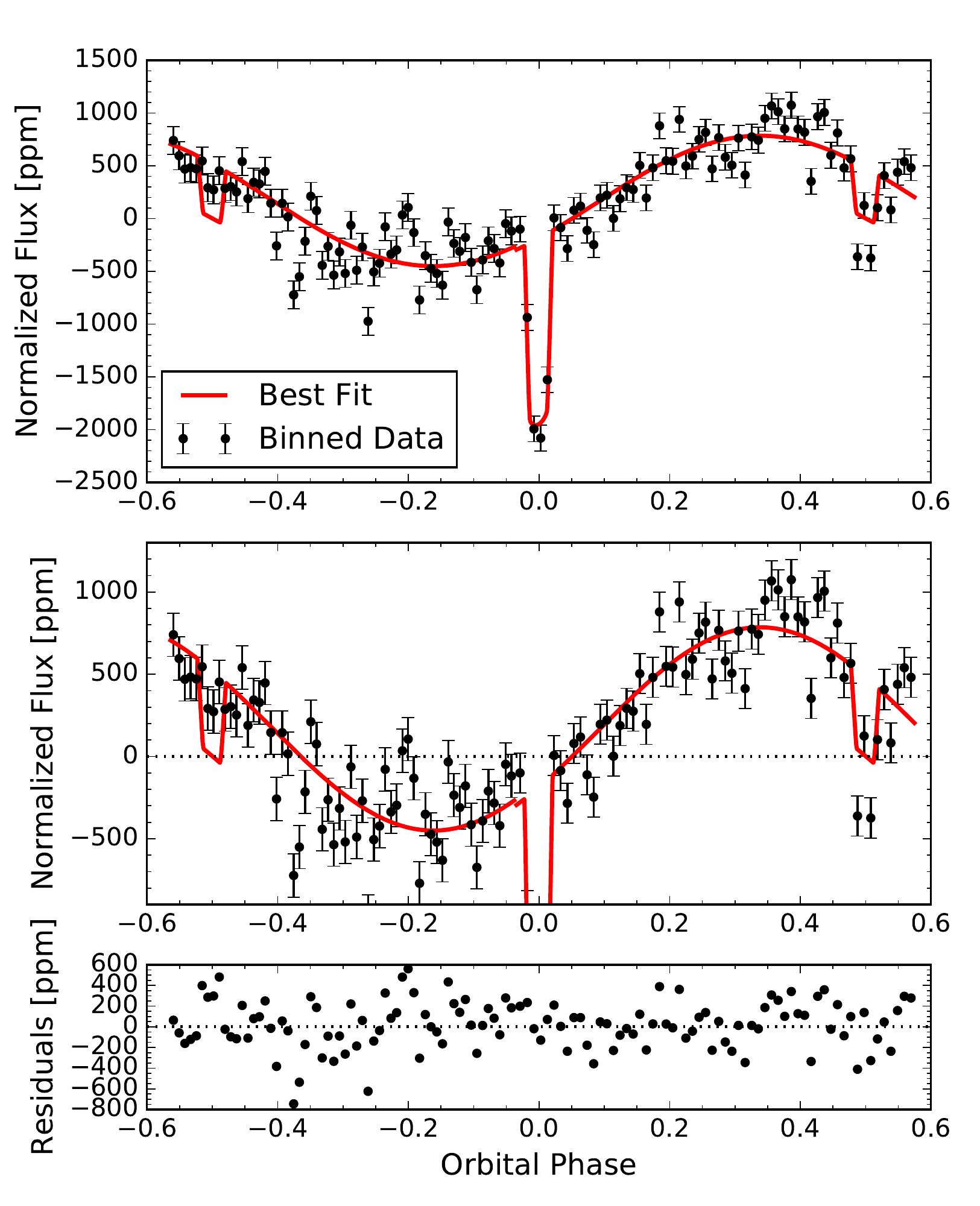}
\caption{Spitzer/IRAC1 (3.6\,\micron) observations.  From top to
  bottom: the full set of observations, the full data set again (but
  stretched to highlight the phase variations), and the residuals to
  the fit. The binned photometry is indicated by the black points, and
  the red line is the best-fit model. The best-fit model is
  nonphysical, requiring negative night-side flux.  This model reaches
  only 2.3$\times$ the photon-noise limit (worse when restricted to
  physically plausible models) and the residuals exhibit considerable
  correlated noise (Fig.~\ref{fig:bindown1}).
 \label{fig:phase1} }
\end{figure}

\begin{figure}[b!]
\includegraphics[width = 0.92\textwidth]{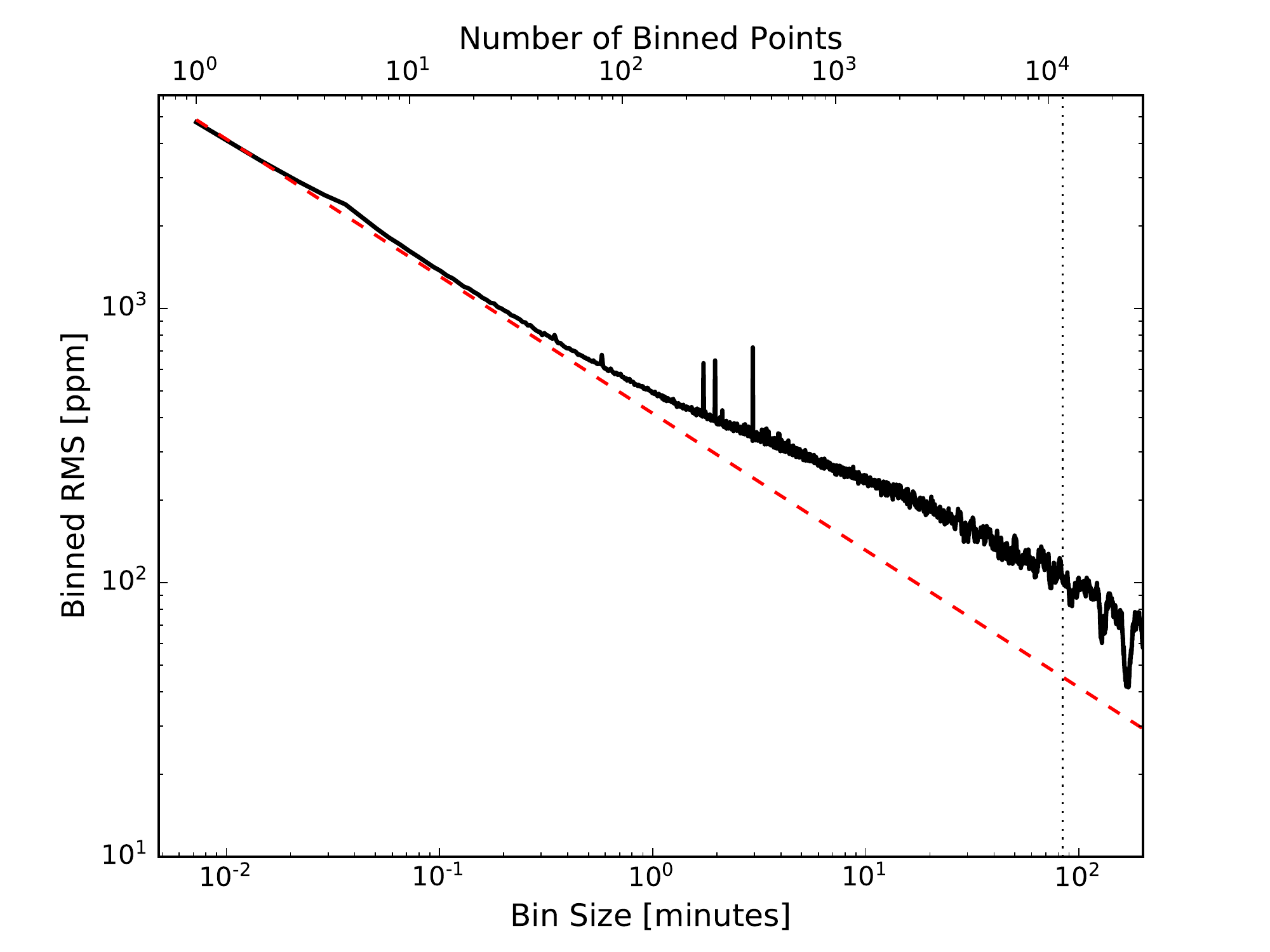}
\caption{The dispersion of the binned residuals (solid line) to the
  3.6\,\micron\ light curve (Fig.~\ref{fig:phase1}) shows  significant evidence for
  correlated noise. The dashed line shows the expectation for wholly
  uncorrelated errors, which scale as $N^{-1/2}$, and the vertical
  dotted line indicates the transit duration.
   \label{fig:bindown1}}
\end{figure}

\begin{figure}[b!]
\includegraphics[width = 0.92\textwidth]{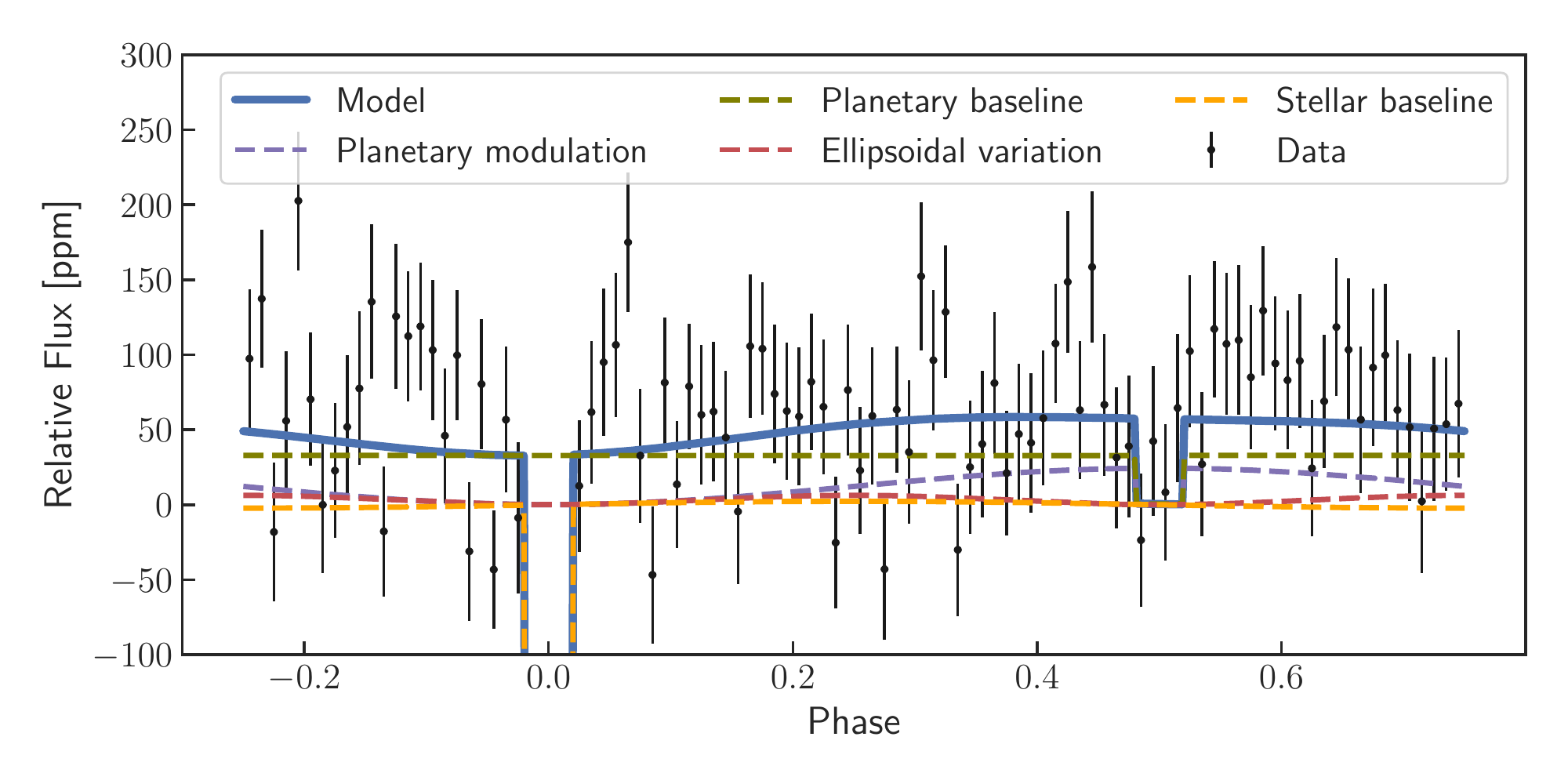}
\includegraphics[width = 0.92\textwidth]{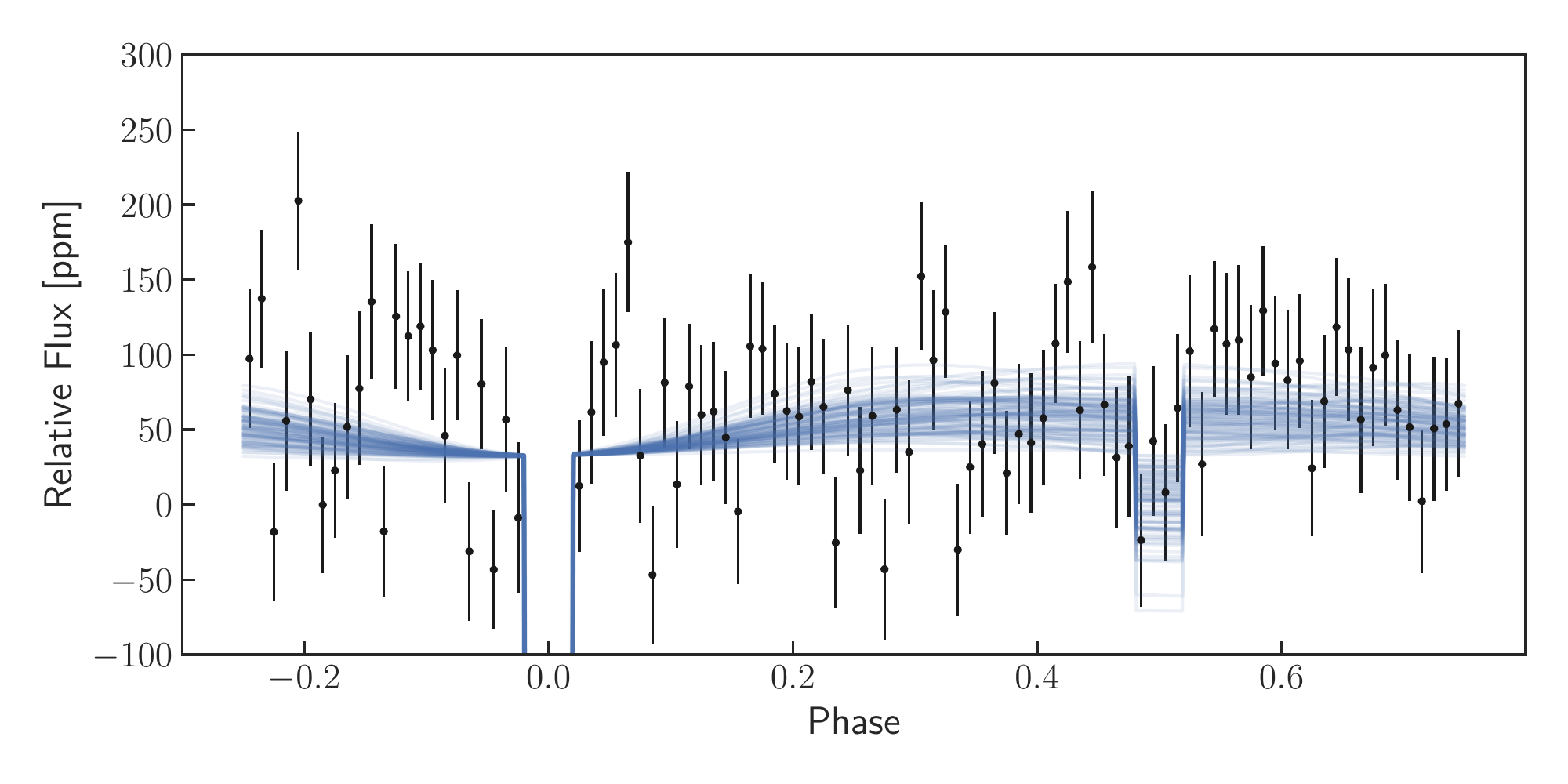}
\caption{{\em TESS} optical-wavelength photometry of LTT~9779, shown
  by the black points and error bars. {\em Top:} The best-fit total
  model (blue) along with individual components. {\em Bottom:}
  Posterior distribution from our modeling analysis, showing a likely
  detection of the planet's secondary eclipse but no clear detection
  of the planet's phase variation.  \label{fig:tessphase}}
\end{figure}

\begin{figure}[b!]
\includegraphics[width = 0.92\textwidth]{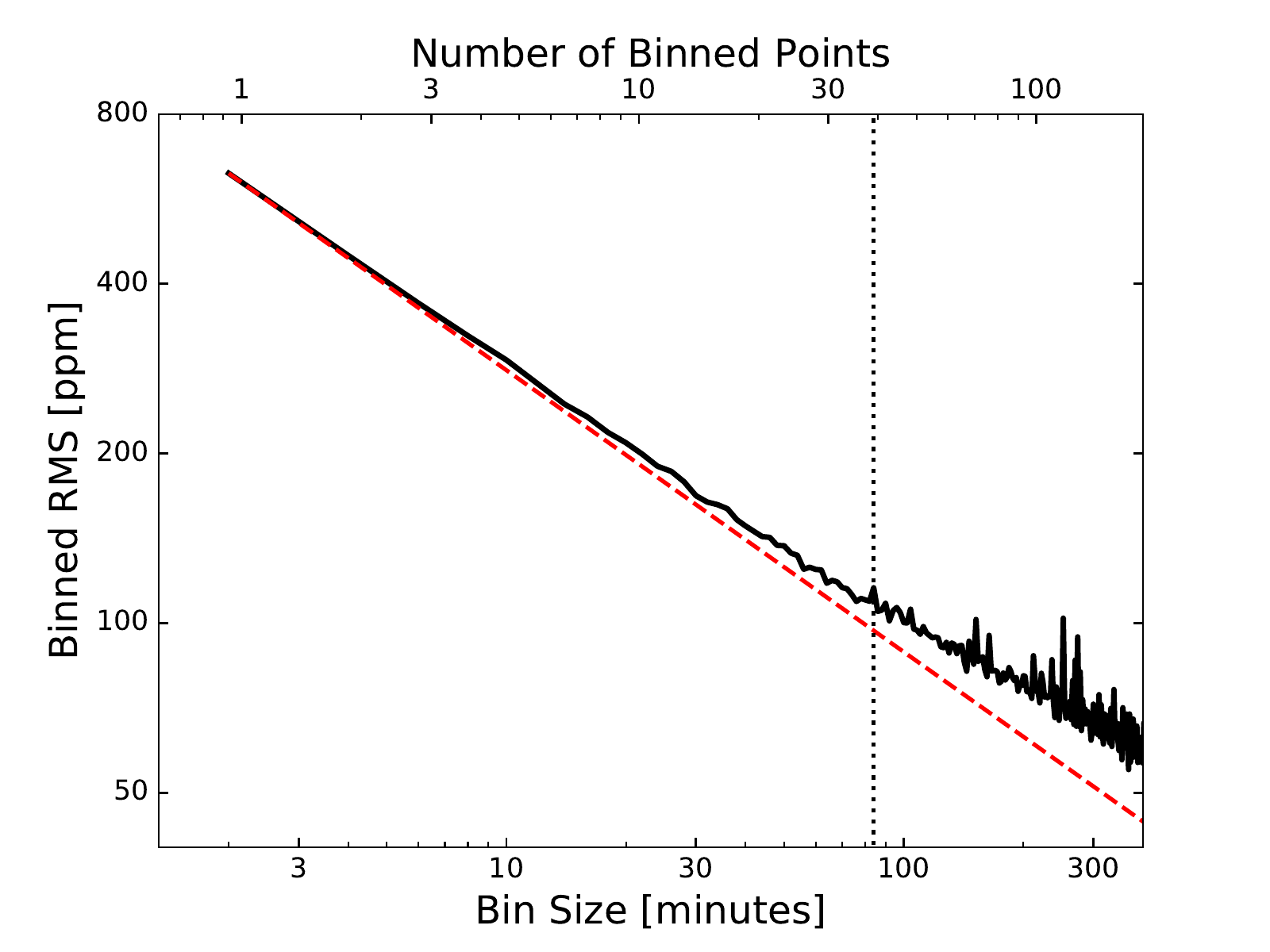}
\caption{Same as Figs.~\ref{fig:bindown2} and~\ref{fig:bindown1}, but
  for the {\em TESS} data shown in Fig.~\ref{fig:tessphase}.
  \label{fig:bindowntess}}
\end{figure}

We have presented infrared and optical phase curve observations of
this unusual hot Neptune, building on our analysis of its day-side
emission spectrum through secondary eclipses in
\cite{dragomir:2020}. The planet's phase curve is clearly seen in our
Spitzer 4.5\,\micron\ photometry (Fig.~\ref{fig:phase2}); these data
detect the signature of heat recirculation on this planet, with a
$3\sigma$ confidence interval on the day-night temperature contrast of
$\deltat\pm\uthreedeltat$~K.  Our observations plug a glaring gap,
since Fig.~\ref{fig:phasecurvetargets} shows that no infrared phase
curve data exist for any similar planet (i..e, within a factor of
three of \planet 's mass and a factor of two of its temperature).

\begin{figure}[b!]
\includegraphics[width = 0.92\textwidth]{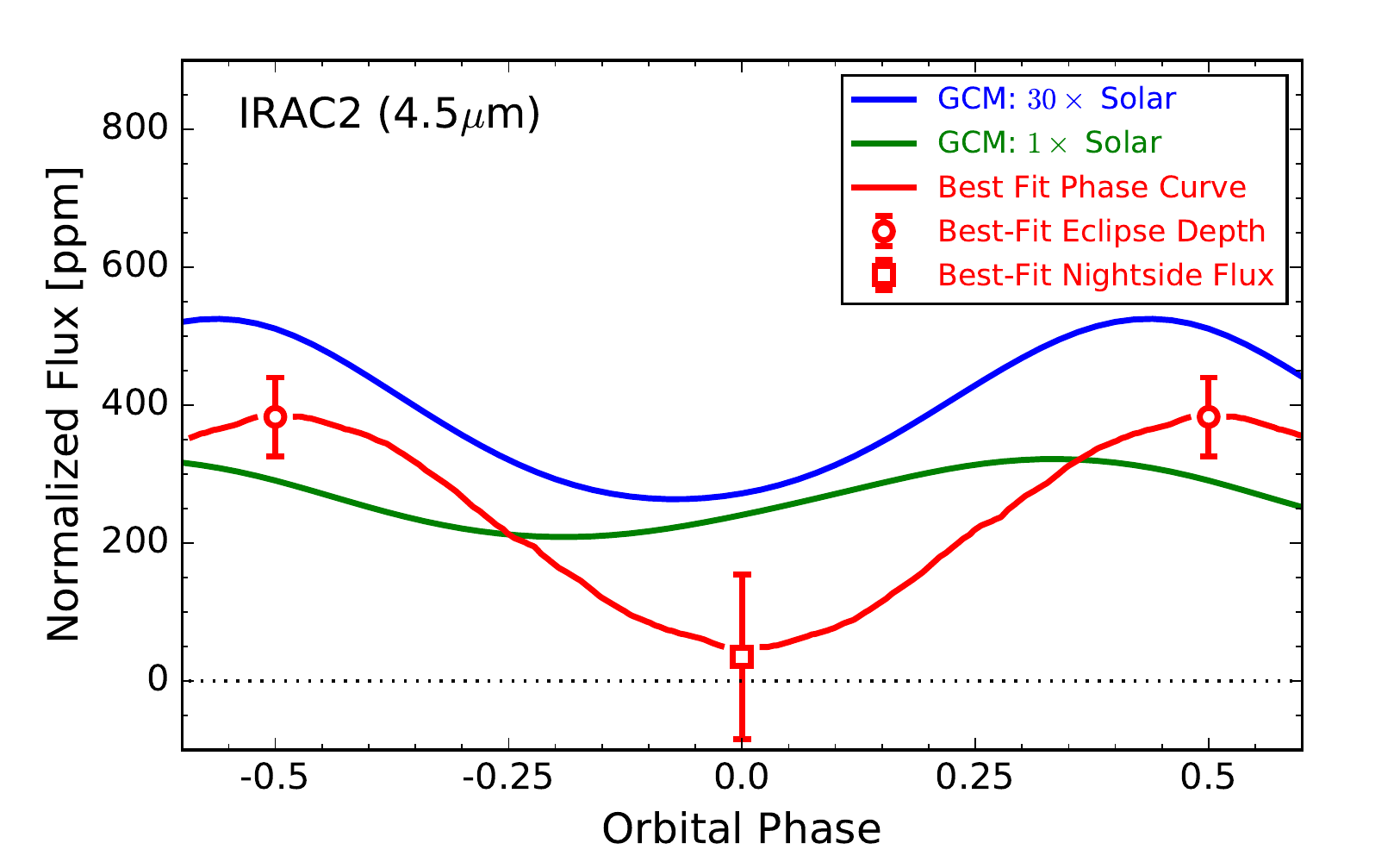}
\includegraphics[width = 0.92\textwidth]{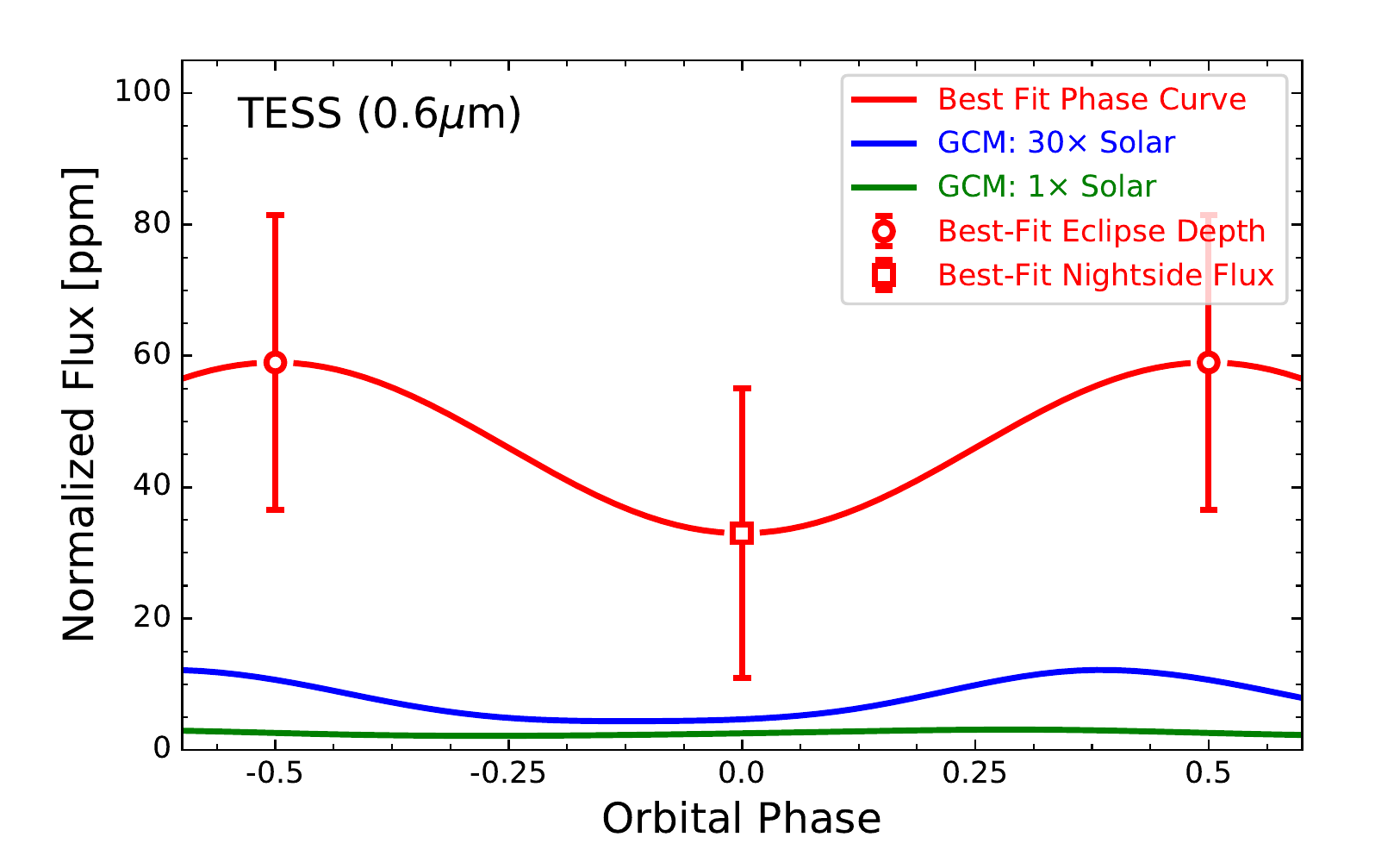}
\caption{Best-fit model to our phase curve data (red curves) along
  with predicted phase curves from our general circulation models
  (blue and green curves). The gaps in the red curve indicate the
  times of secondary eclipse and primary transit; the error bars
  indicate the measured day-side flux (circles) and night-side flux
  (square). Neither GCM closely matches the observations, either at
  4.5\,\micron\ ({\em Spitzer}, top) or at 0.6\,\micron\ ({\em TESS},
  bottom).   \label{fig:gcmcomparison}}
\end{figure}

\begin{figure}[b!]
\includegraphics[width = 0.75\textwidth]{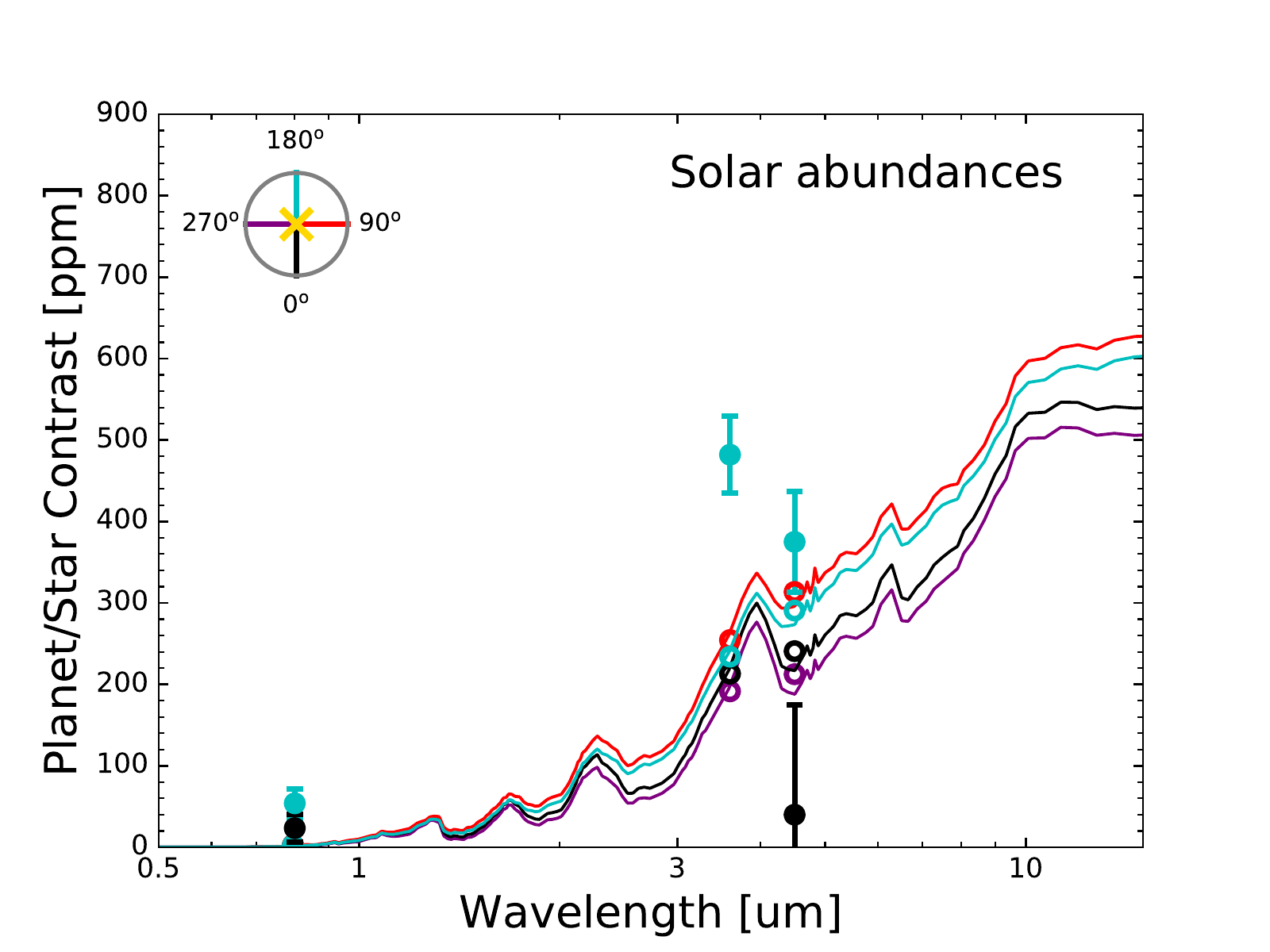}
\includegraphics[width = 0.75\textwidth]{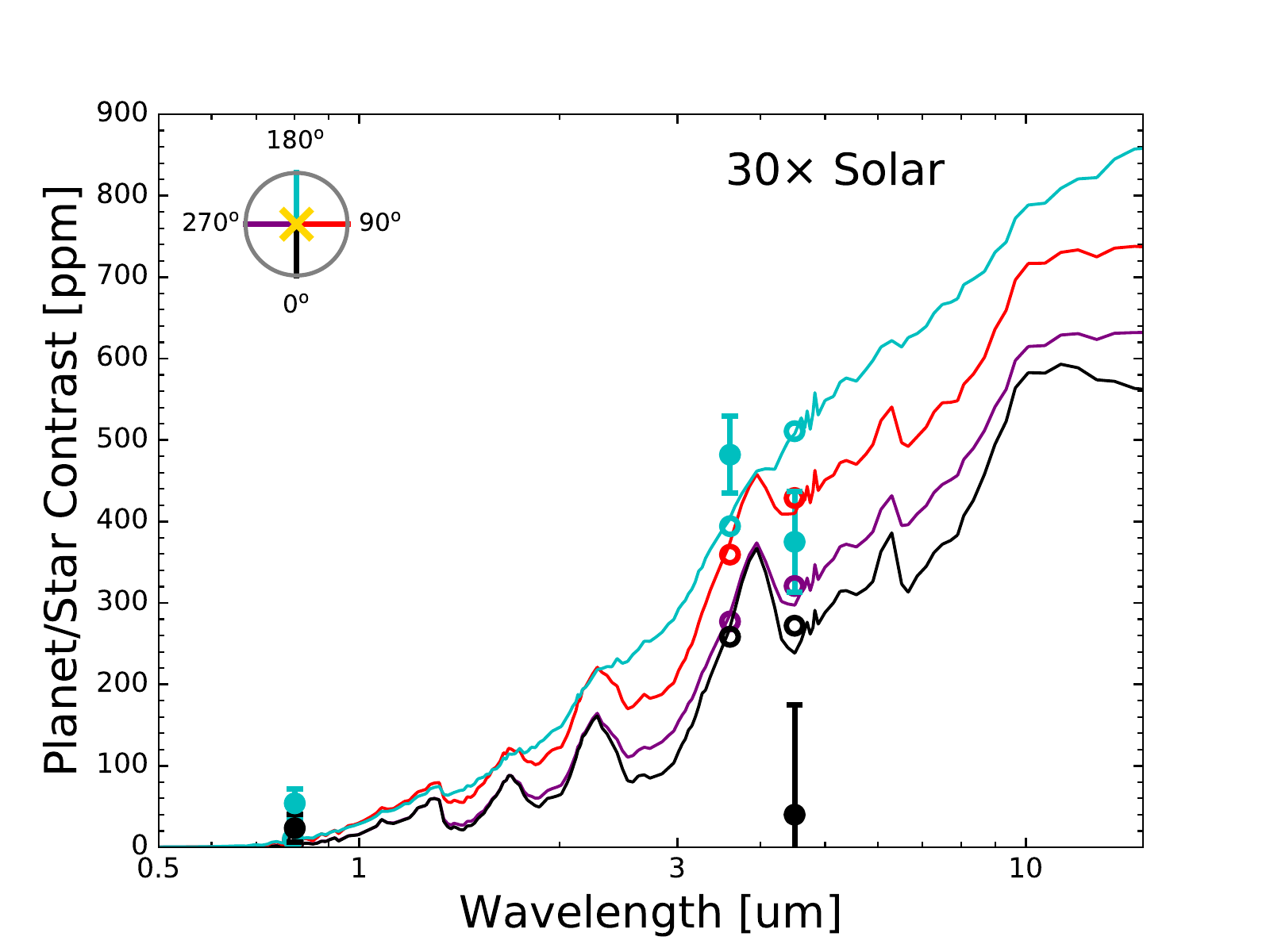}
\caption{Planet/star contrast from our GCM simulations of LTT~9779b,
  assuming Solar abundances ({\em left}) and 30$\times$ Solar
  abundances ({\em right}), at four orbital phases.  Black, night-side,
  as seen during transit (the planet's night side); red, 90$^{\rm o}$
  after transit; light blue, as seen during secondary eclipse (day
  side); and purple, 90$^{\rm o}$ after secondary eclipse. The solid
  points with error bars show the TESS and Spitzer measurements of the
  day- and night-side fluxes; the open circles indicate the
  band-averaged model points. The key in the top left corner is
  color-coded with the spectra to illustrate the
  sequence. \label{fig:phasecontrast}}
\end{figure}

\begin{figure}[b!]
\includegraphics[width = 0.49\textwidth]{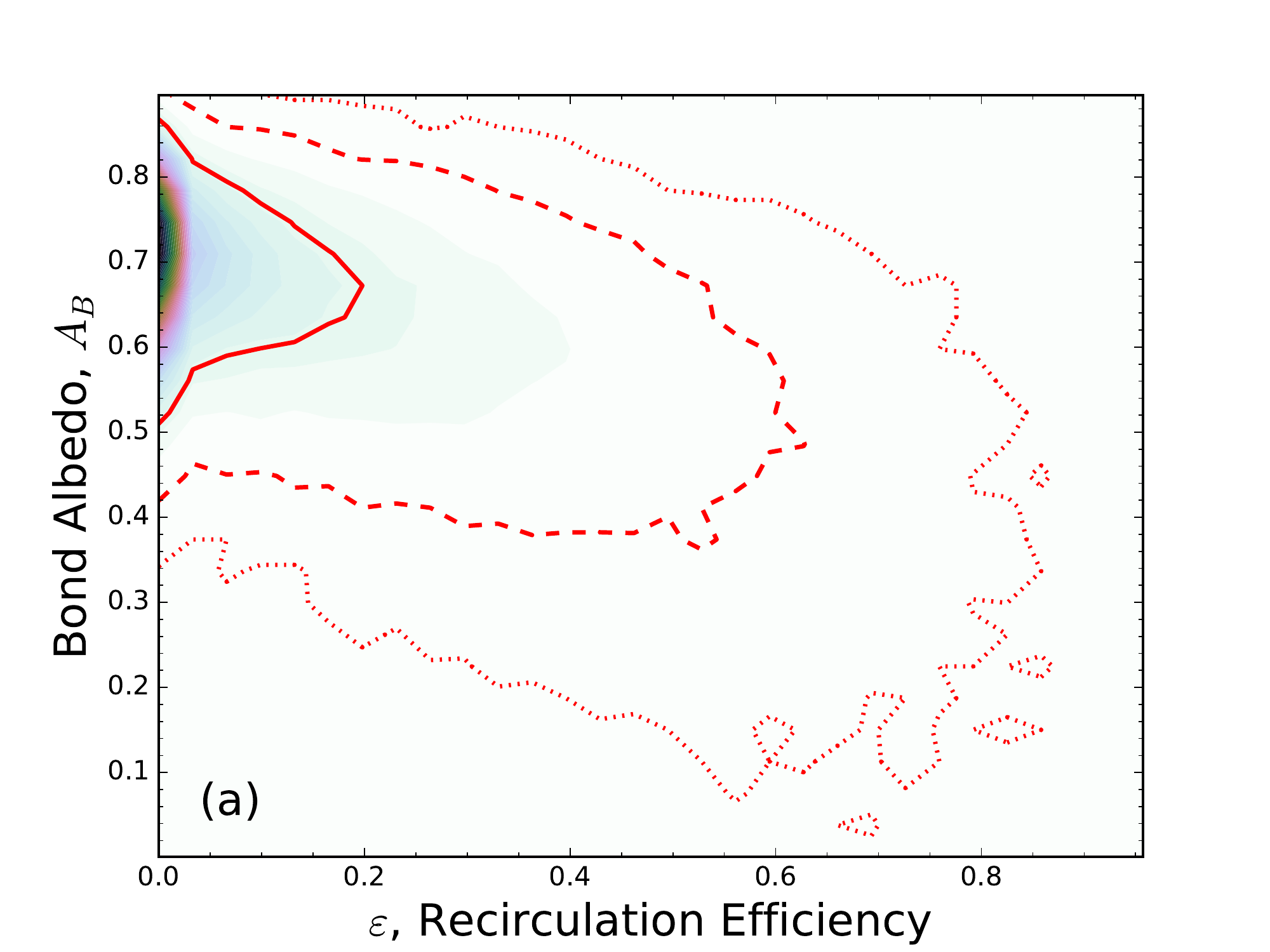}
\includegraphics[width = 0.49\textwidth]{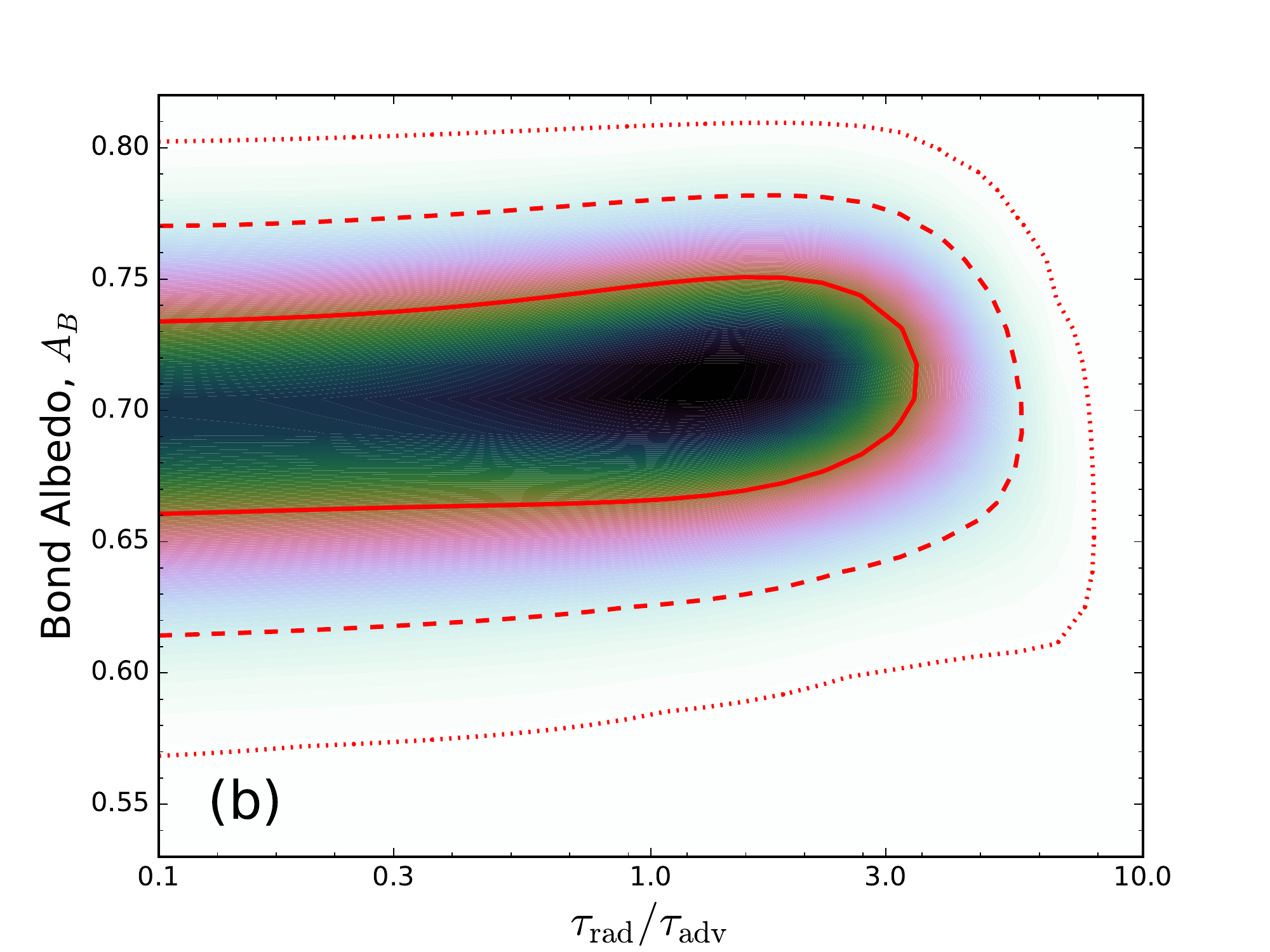}
\caption{Probability distribution of atmospheric parameters derived
  from our 4.5\,\micron\ phase curve, with red lines indicating
  confidence intervals: 1$\sigma$ (solid), 2$\sigma$ (dashed), and
  3$\sigma$ (dotted).  Panel (a) shows Bond Albedo and energy
  recirculation efficiency, while panel (b) shows Bond Albedo and the
  timescale ratio
  $\tau_\mathrm{rad}/\tau_\mathrm{adv}$.  \label{fig:albedoposteriors} }
\end{figure}

\begin{figure}[b!]
  \centering
  \includegraphics[width = 0.31\textwidth]{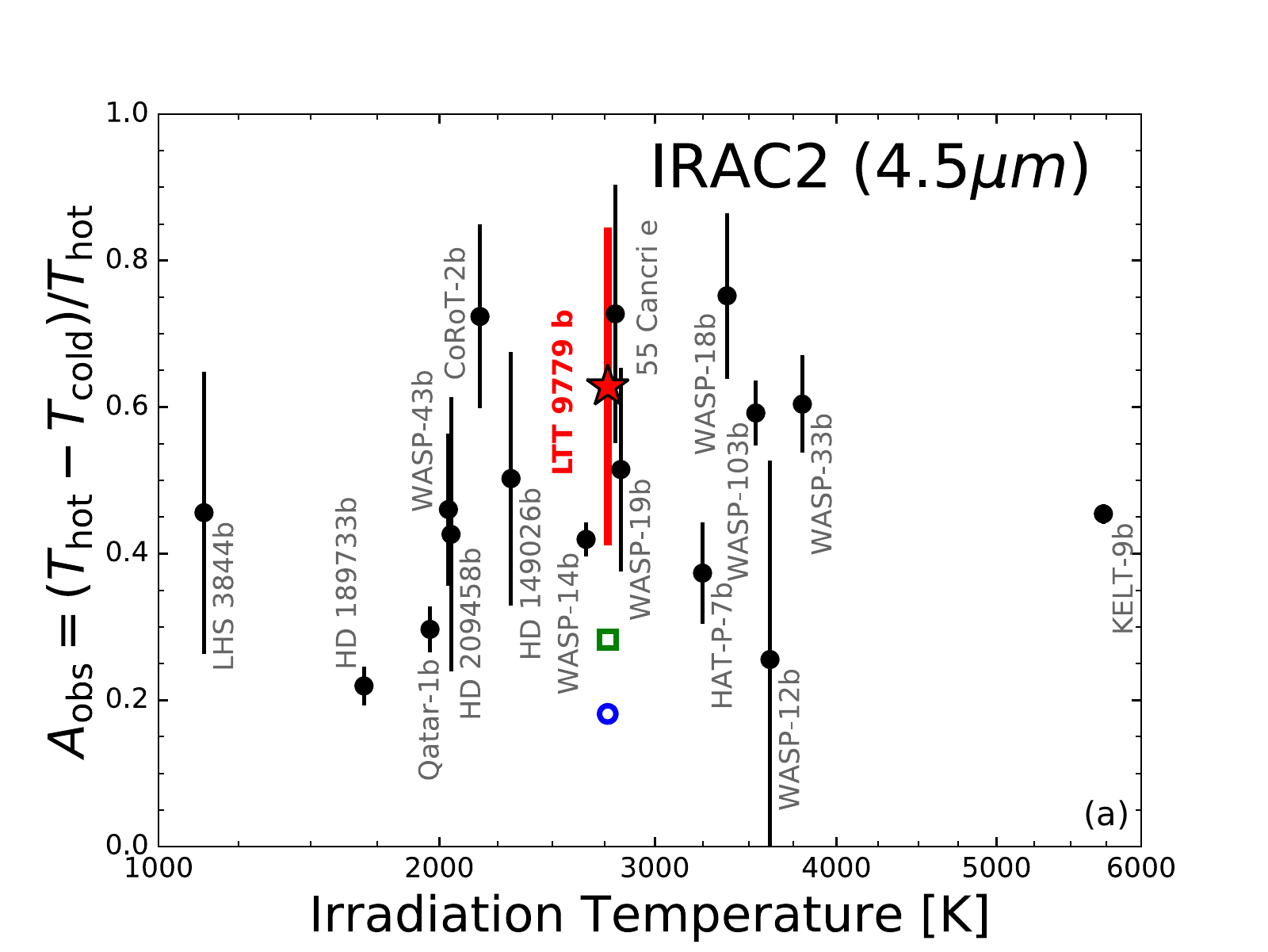}
\includegraphics[width = 0.31\textwidth]{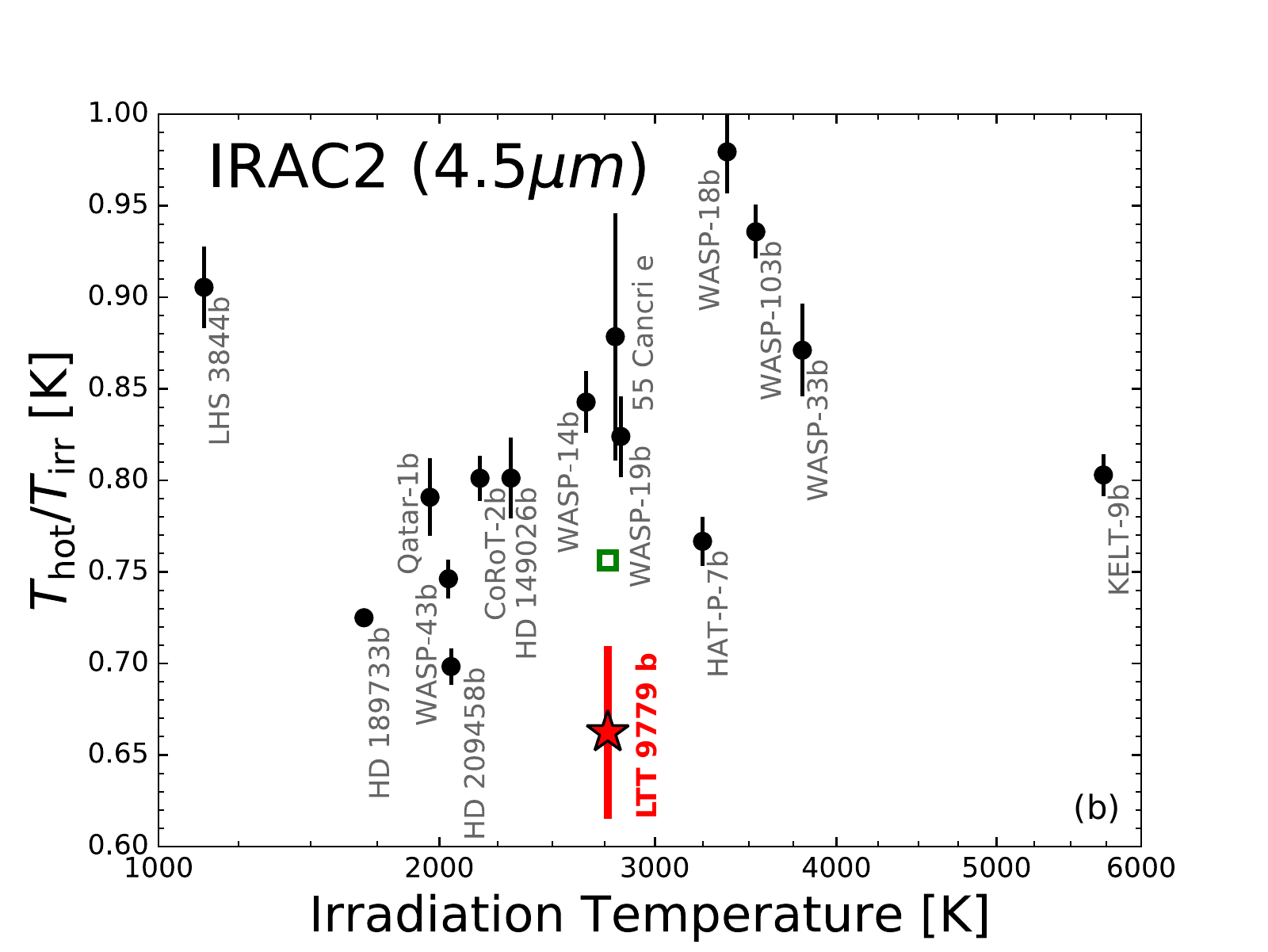}
\includegraphics[width = 0.31\textwidth]{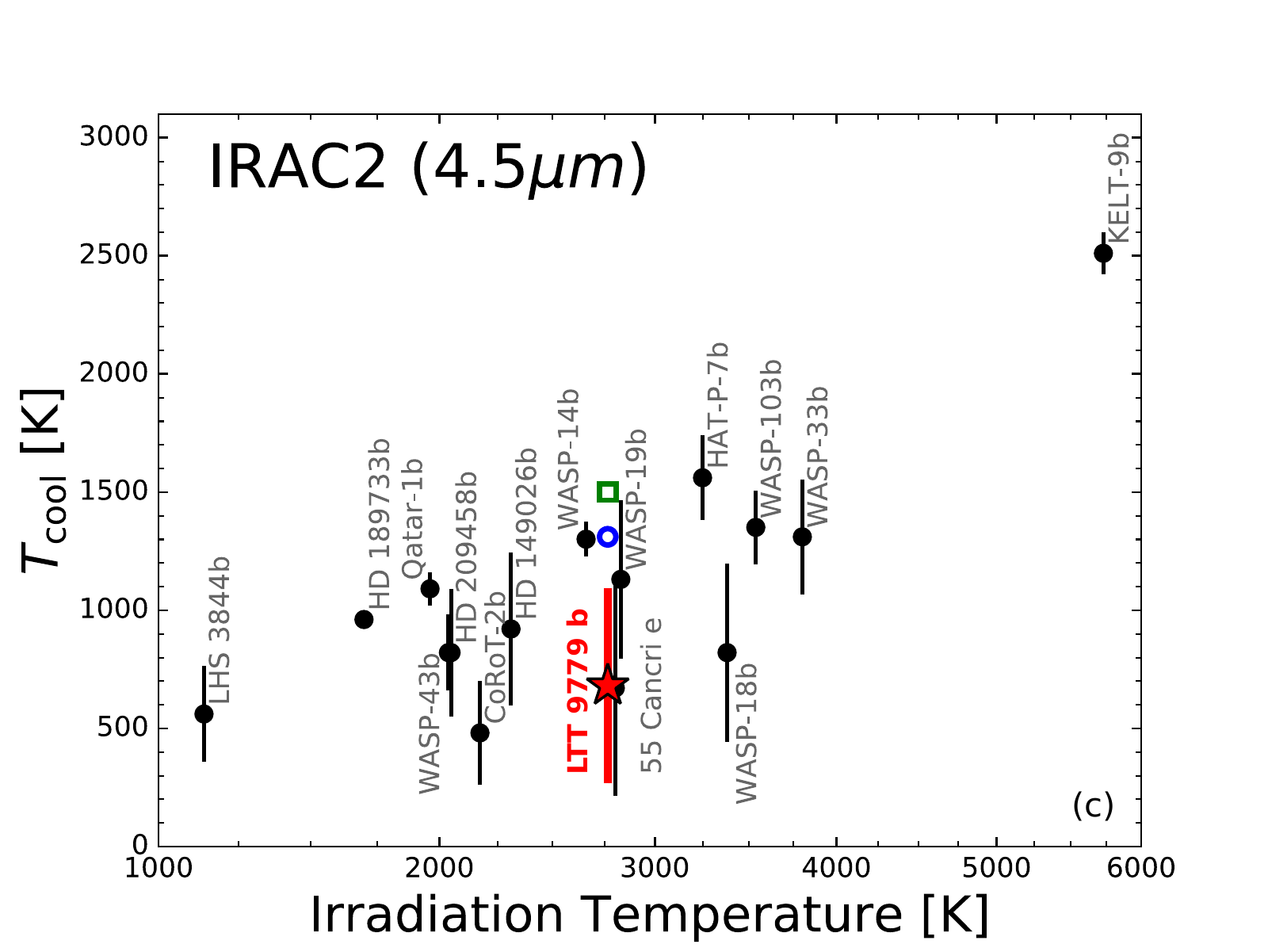}\\
\includegraphics[width = 0.31\textwidth]{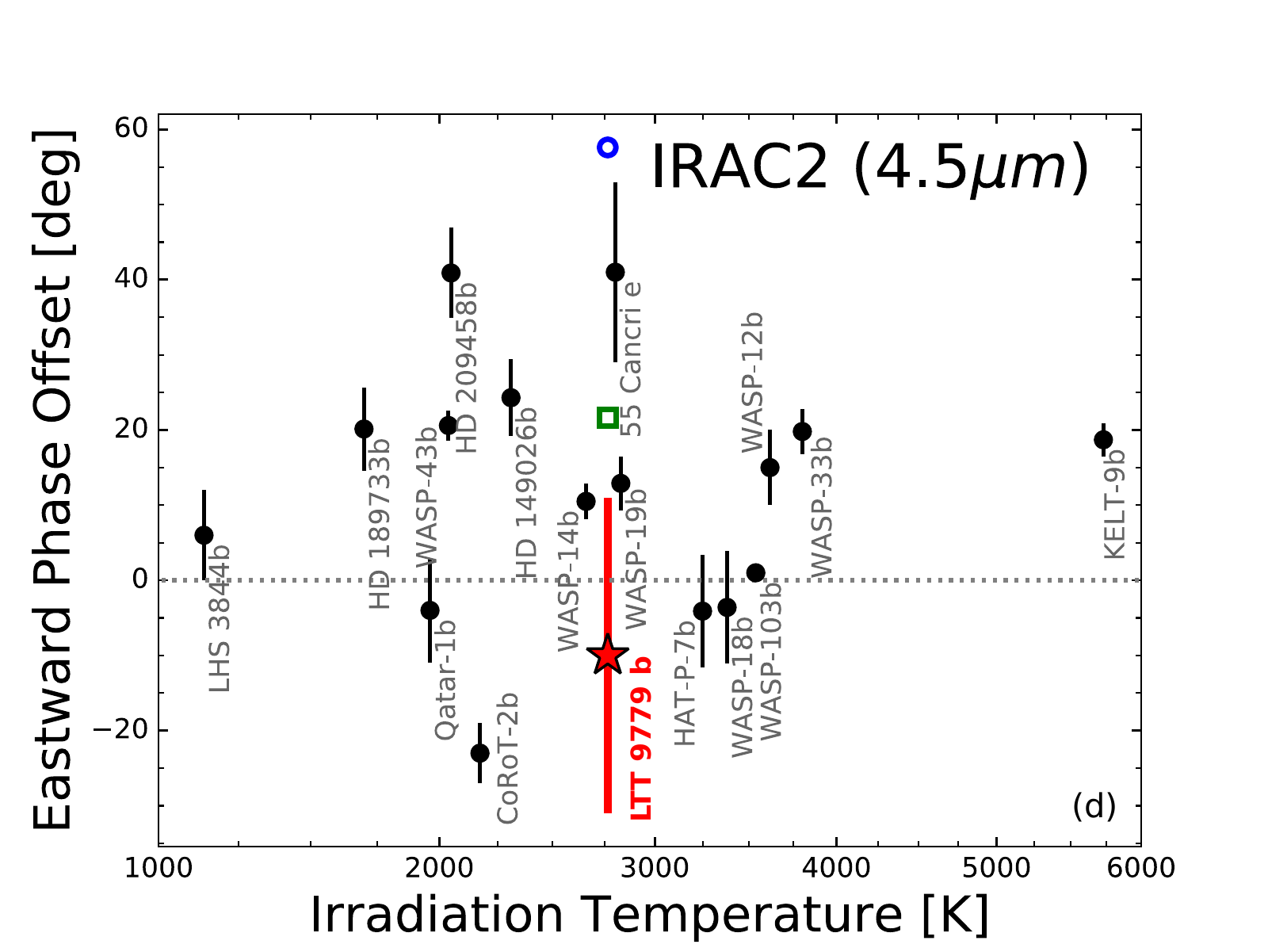}
\includegraphics[width = 0.31\textwidth]{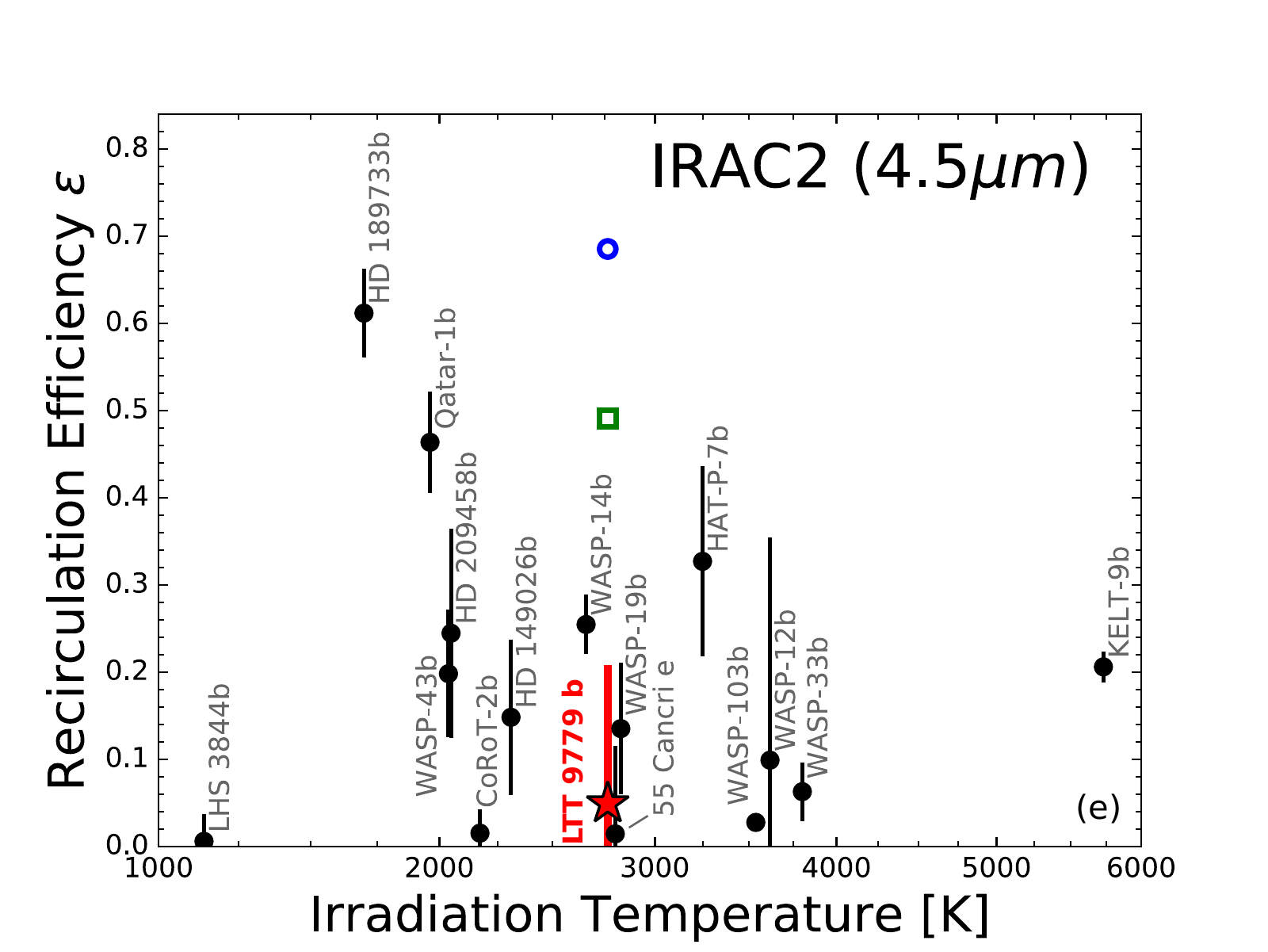}
\includegraphics[width = 0.31\textwidth]{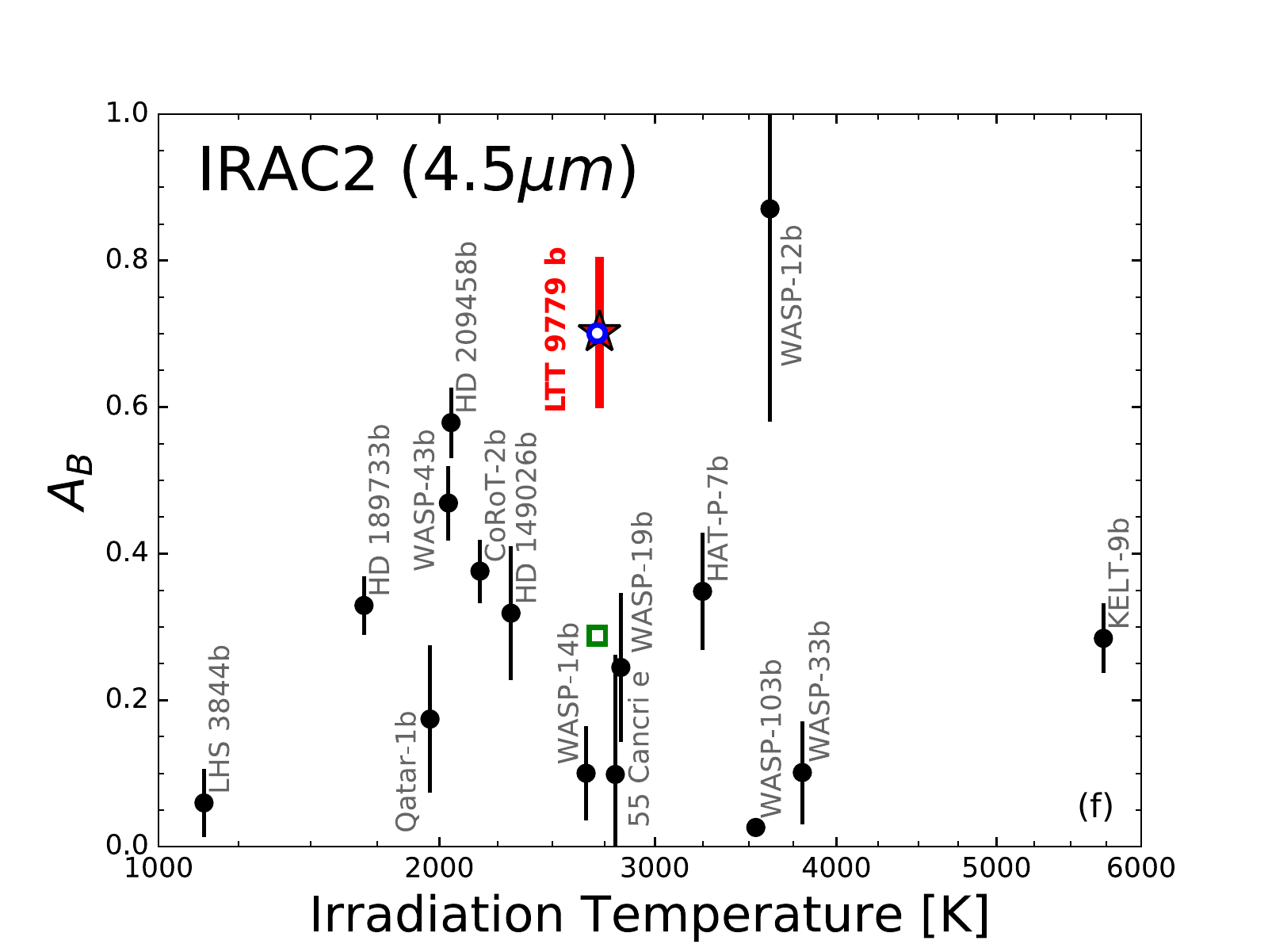}\\
\includegraphics[width = 0.31\textwidth]{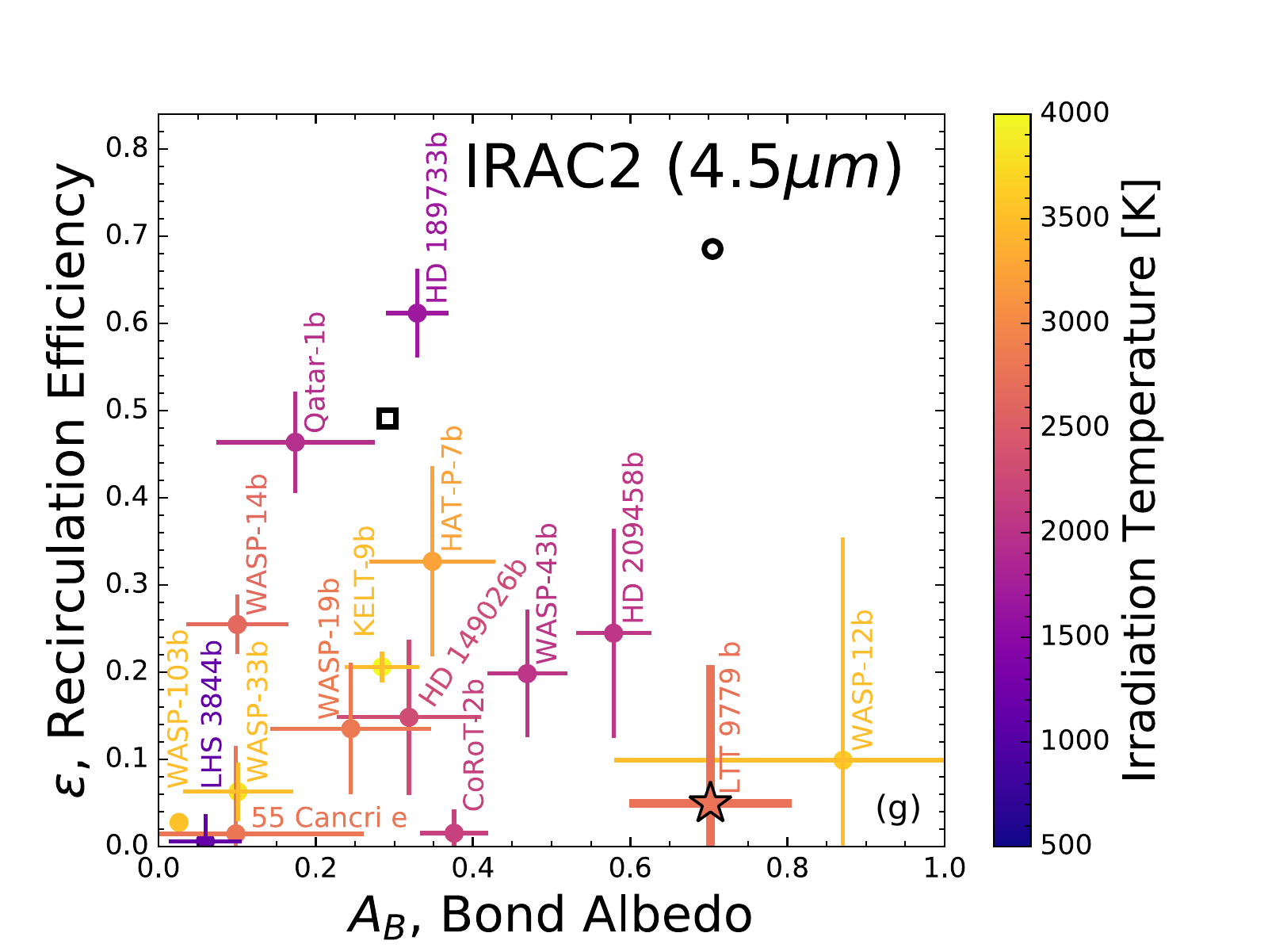}
\includegraphics[width = 0.31\textwidth]{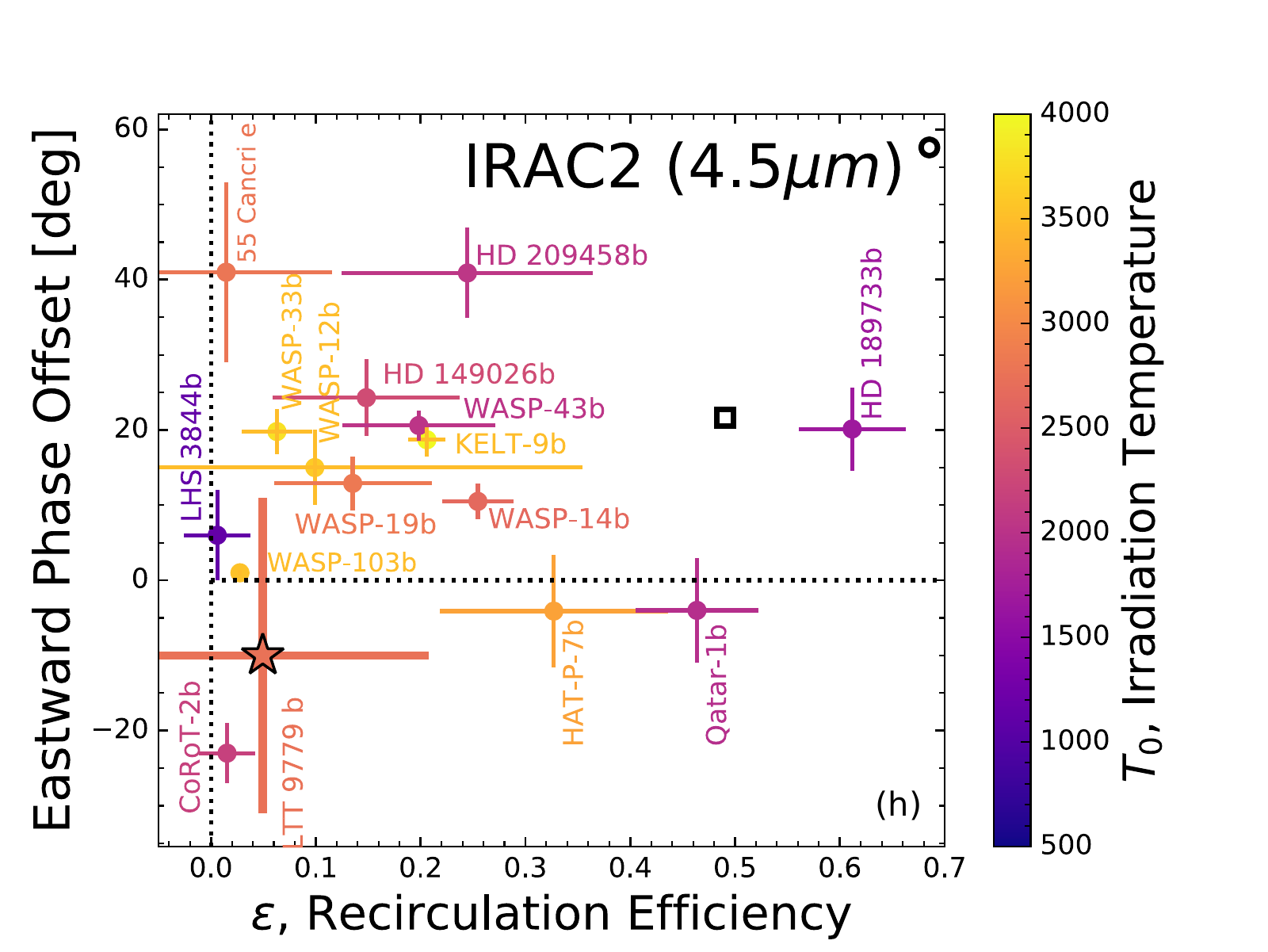}
\includegraphics[width = 0.31\textwidth]{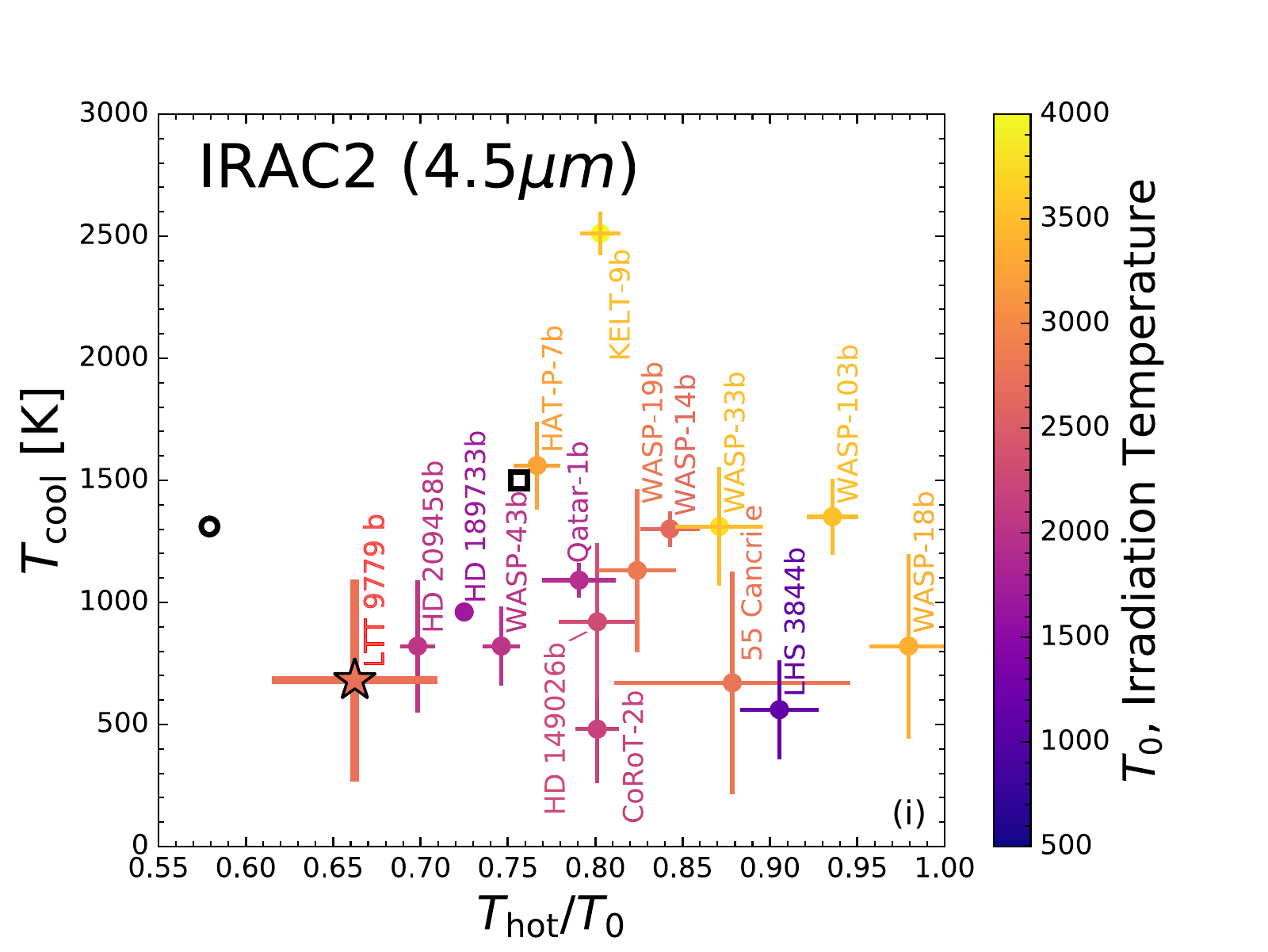}
\caption{Derived parameters from all IRAC 4.5\,\micron\ phase curve
  measurements of planets on nearly-circular orbits.  LTT~9779b is
  indicated by the red star, while our GCM predictions are
  indicated by the open blue circle (Solar metallicity) and green
  square (30$\times$ Solar). \label{fig:phasecurvesample}  }
\end{figure}

\begin{figure}[b!]
\includegraphics[width = 0.92\textwidth]{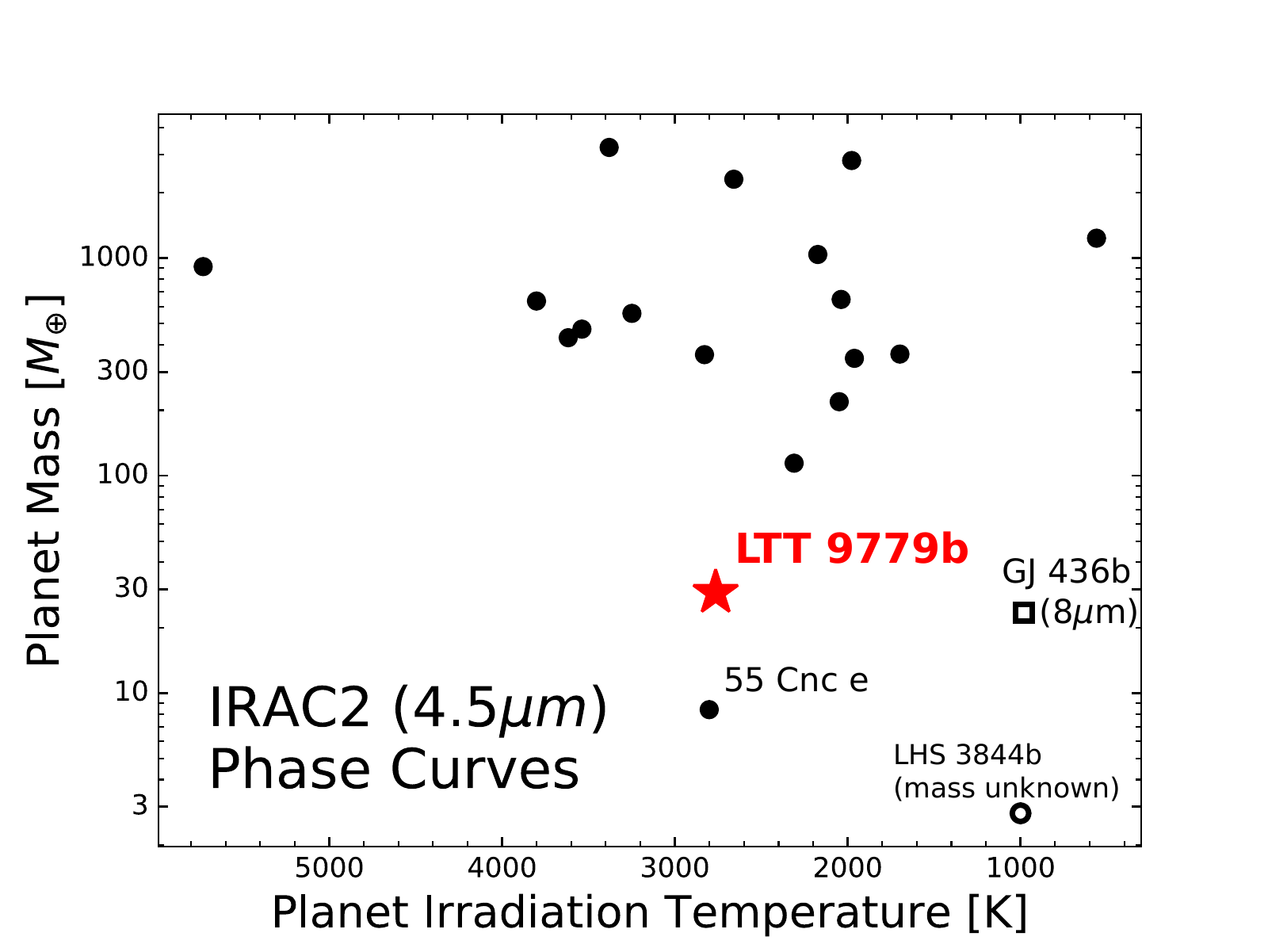}
\caption{All planets with published Spitzer/IRAC2 (4.5\,\micron) phase
  curves.  Only four planets with $M_p \lesssim 30 M_\oplus$ have
  measured infrared phase curves: 55~Cnc~e, LHS 3844b (whose mass is
  unknown), and GJ 436b (at 8\,\micron). \label{fig:phasecurvetargets}}
\end{figure}

Overall, via several lines of evidence our measurements hint at an
atmospheric metallicity enhanced above the Solar level. These
arguments are mainly derived from the strong constraints placed on the
planet's global circulation patterns from the 4.5\,\micron\ phase
curve. First, the {\update eastward} phase offset of $\phaseoffset \pm
\uphaseoffset^{\rm{o}}$ (to the west) is more consistent with that
predicted by our general circulation models (GCMs) that are
preferentially enhanced in heavy elements (see
Fig.~\ref{fig:gcmcomparison} and Table~\ref{tab:gcmcomparison}). Such
a small phase offset also indicates a low heat recirculation
efficiency and relatively small ratio of radiative to advective
timescales, consistent with the values seen for hot Jupiters at
comparable temperatures.  In addition, LTT~9779b's 4.5\,\micron\ phase
curve amplitude of $\phaseamplitude \pm \uphaseamplitude$~ppm is also
consistent with our enhanced-metallicity GCM predictions (see
Fig.~\ref{fig:gcmcomparison} and Table~\ref{tab:gcmcomparison}),
though that model's orbit-averaged infrared emission is elevated
compared to our observations; future modeling at higher atmospheric
metallicity and including the effects of aerosols {\update may be
  needed to conclusively interpret our observations}
\citep[cf.\ ][]{kataria:2015,parmentier:2016a,keating:2019}.  {\update
  Including the 3.6\,\micron\ day-side observation in the GCM
  comparison (Fig.~\ref{fig:phasecontrast}) also shows a better
  (though far from perfect) match between the observations and our
  higher-metallicty model. }

Furthermore, the GCM predictions may also under-predict the planet's
optical-wavelength emission seen by TESS, though this result is more
tentative (TESS is scheduled to observe LTT~9779b for another
month-long sector in late 2020, which should tighten the measurements
at optical wavelengths). {\update The optical observations may hint at
  a nonzero geometric albedo, while the two-channel {\em Spitzer}
  measurements do not tightly constrain the planet's Bond albedo.}



Finally, further support for a high-metallicity atmosphere comes from
our comparison of LTT~9779b's thermal phase curve and day-side
emission measurements with two similarly-irradiated hot Jupiters,
WASP-14b and WASP-19b (see Table~\ref{tab:hotplanets}). In particular,
although the surface gravity and irradiation temperature of WASP-19b
differ from those of LTT~9779b by $<$10\%, the two planets have
qualitatively different emission spectra at $>99\%$ confidence.  The
hot Jupiters WASP-14b and WASP-19b both exhibit approximately
featureless, pseudo-blackbody emission spectra devoid of spectral
features, in contrast to the absorption inferred in our hot Neptune's
broadband emission spectrum \citep{dragomir:2020}; {\update models
  also suggest that stronger absorption at this irradiation
  tempreature is consistent with a higher-metallicity atmosphere
  \citep{mansfield:2020ep3}.}

The primary, two-year TESS mission is nearly complete. When done, 85\%
of the sky will have been surveyed for planets transiting nearby
stars.  To date, LTT~9779b remains one of the best targets of its type
-- i.e., among highly irradiated hot Neptunes that are highly
favorable for thermal emission measurements -- and so it is likely to
remain one of the most easily-characterizable exoplanets in its class.
Future phase curve and eclipse observations of \planet\ and other
similar planets will provide the impetus for the next generation of
global circulation modeling of Neptune-size planets, supporting
similar observations of many such objects with JWST, ARIEL, and future
observatories.

\acknowledgments {\update The authors thank our anonymous referee in
  advance, for constructive comments that improved the quality of this
  work.  We also thank Drs.\ K.~Stevenson and P.~Cubillos for their
  advice and insights into the \texttt{POET} code.}

I.J.M.C. acknowledges support from the NSF through grant AST-1824644,
and from NASA through Caltech/JPL grant RSA-1610091.  {\update
  D.D. gratefully acknowledges support for this work from NASA through
  Caltech/JPL grant RSA-1006130}.  T.D. acknowledges support from
MIT's Kavli Institute as a Kavli Postdoctoral Fellow. J.S.J.
acknowledges support by FONDECYT grant 1201371 and partial support
from CONICYT project Basal AFB-170002.

We acknowledge the use of \TESS\ Alert data, which was in a beta test
phase, from pipelines at the \TESS\ Science Office and at the
\TESS\ Science Processing Operations Center. This research has made
use of the Exoplanet Follow-up Observation Program website, which is
operated by the California Institute of Technology, under contract
with the National Aeronautics and Space Administration under the
Exoplanet Exploration Program. This paper includes data collected by
the \TESS\ mission, which are publicly available from the Multimission
Archive for Space Telescopes (MAST).  Resources supporting this work
were provided by the NASA High-End Computing (HEC) Program through the
NASA Advanced Supercomputing (NAS) Division at Ames Research Center
for the production of the SPOC data products.



 \facility{TESS, Spitzer }

 \cite{may:2020}
 \bibliographystyle{aasjournal}

\clearpage

\appendix \cite{cowan:2011circ} present a simple two-parameter model
for atmospheric phase curves, in which a planet's global temperature
distribution is determined by the interplay between its Bond albedo,
$A_B$, and the ratio of its radiative and advective timescales,
$\varepsilon' \equiv \tau_\mathrm{rad}/\tau_\mathrm{adv}$. We present here a further
analytic approximation to this modeling framework which may be
pedagogically useful.

This model assumes a coordinate system in which $\theta=0$ at the
North pole and $\pi$ at the South pole, while $\phi=0$ at the
substellar longitude, $-\pi/2$ at dawn, and $\pi/2$ at sunset.  It
defines planetary temperature $T$ in terms of $T' = T/\tirr$, where
$\tirr$ is a fiducial planetary temperature:
\begin{equation}
  \tirr \equiv T_\mathrm{eff} (1 - A_B)^{1/4} \sin^{1/4} \theta \left( \frac{R_*}{a} \right)^{1/2} .
\end{equation}
Subject to energy balance, the temperature of gas parcels all around
the planet are then the solutions to the differential equation
\begin{equation}
\frac{d T'}{d \phi} = \frac{2 \pi \tau_\mathrm{adv}}{\tau_\mathrm{rad}} \left( \max(\cos \phi, 0) - {T'}^4 \right) .
\label{eq:parceltemp3}
\end{equation}

\cite{cowan:2011circ} provide an analytic solution to
Eq.~\ref{eq:parceltemp3} for the planet's night-side but numerical
solutions are required for the day side. The analytic solution for the
night-side temperature can be found by integrating from dusk (when the
parcel stops absorbing energy, $\phi=\pi/2$) until some later phase
$\phi$ (up until dawn, $\phi=3 \pi/2$), and its solution is
\begin{equation}
  T'_\mathrm{night}(\phi) = \left( 6 \pi \frac{ \tau_\mathrm{adv}}{\tau_\mathrm{rad}} \left[\phi - \frac{\pi}{2}\right] + (T'_\mathrm{dusk})^{-3} \right)^{-1/3} .
\label{eq:transportnight}
\end{equation}

On the planet's day side, a second-order approximation can provide
some additional insights hidden by the more exact (but
numerically-derived) solution.  By assuming that $T'$ is a quadratic
function of $\phi$ and expanding $\cos \phi \approx 1 - \phi^2/2$, one
obtains
\begin{equation}
T'_\mathrm{day} \approx \left( 1 - \frac{\varepsilon'^2}{64 \pi^2} \right) + \frac{\varepsilon'}{32 \pi}  \phi - \frac{1}{8} \phi^2 .
\label{eq:transport2ndorder}
\end{equation}
This approximate analytic solution shows that the temperature of
planetary gas reaches a maximum temperature at longitude
\begin{equation}
  \phi_\mathrm{max} \approx \frac{1}{8 \pi} \frac{\tau_\mathrm{rad}}{\tau_\mathrm{adv}} .
\end{equation}
By setting $\phi=\phi_\mathrm{max}$ in Eq.~\ref{eq:transport2ndorder},
we obtain the temperature of the day-side hot spot:
\begin{equation}
T'_\mathrm{day,max}  = T'_\mathrm{day}(\phi_\mathrm{max}) \approx 1 - \frac{7 \varepsilon'^2}{512 \pi^2} .
\label{eq:tdaymax}
\end{equation}

\begin{figure}[b!]
\begin{center}
\includegraphics[width = 0.92\textwidth]{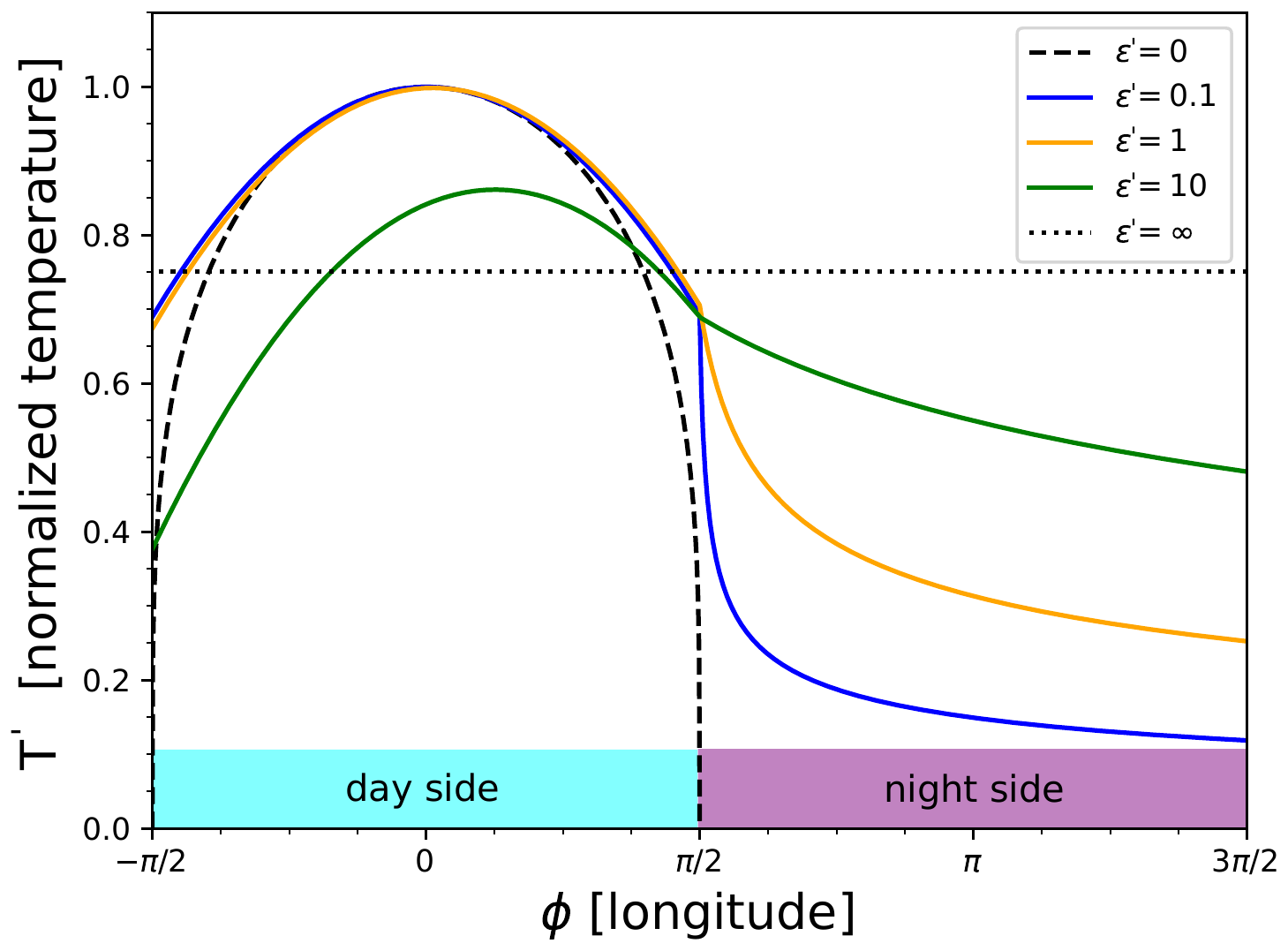}
\end{center}
\caption{\update Planetary temperature vs.\ longitude for the simple
  energy transport model described in the appendix, for different
  values of $\varepsilon' = \tau_\mathrm{rad}/\tau_\mathrm{adv}$. The
  substellar longitude is at $\phi=0$. The solid curves on the day
  side are the second-order analytic, approximate solutions of
  Eq.~\ref{eq:transport2ndorder}, while on the night side are plotted
  the exact solutions of Eq.~\ref{eq:transportnight}. The broken black
  curves are the exact solutions for the limiting cases indicated,
  i.e.\ atmospheres dominated by radiation (dashed) and advection
  (dotted).
   \label{fig:thermalbalance}}
\end{figure}

{\update As shown in Fig.~\ref{fig:thermalbalance},} these approximate
quadratic solutions are a decent match to the exact analytic solution
for low $\tau_\mathrm{rad}/\tau_\mathrm{adv}$, though over-predicting
somewhat the temperature at dawn and dusk.  The approximation breaks
down for $\tau_\mathrm{rad}/\tau_\mathrm{adv} \gtrsim 20$, when
Eq.~\ref{eq:tdaymax} becomes negative, but in this case the phase
curve is nearly flat anyway.

\end{document}